\documentclass[journal]{new-aiaa}
\usepackage[utf8]{inputenc}
\usepackage{textcomp}
\usepackage{subfig}
\usepackage{graphicx}
\usepackage{amsmath}
\usepackage[version=4]{mhchem}
\usepackage{longtable,tabularx}
\usepackage[usenames]{color}
\title{Short-Term Space Occupancy and Conjunction Filter}

\author{Ana S. Rivero \footnote{Ph.D. Candidate, Aerospace Engineering Department, asanchez8@us.es.}}\affil{University of Seville, 41092 Seville, Spain}
\author{Claudio Bombardelli\footnote{Associate Professor, Space Dynamics Group, claudio.bombardelli@gmail.com}}\affil{Technical University of Madrid (UPM), 28040 Madrid, Spain}
\author{Rafael Vazquez\footnote{Professor, Aerospace Engineering Department, rvazquez1@us.es.}}\affil{University of Seville, 41092 Seville, Spain}

\begin{document}

\maketitle

\begin{abstract}
\noindent Conjunction analysis (CA) for resident space objects (RSOs)
is essential for preventing collisions in an increasingly crowded
orbital environment and preserving the operational integrity of satellites.
A first and fundamental step in the CA process is to estimate the
range of altitudes that each object can occupy, throughout an operational
screening time of, typically, a few days. In this paper, a method
is poroposed to analytically evaluate such range of altitudes in a
zonal problem model and for a time horizon of generic duration thereby
generalizing the concept of space occupancy (SO) introduced in a recent
work. The proposed method is exploited to construct a new conjunction
filter that considerably improves the classical apogee-perigee filter
routinely employed in CA. The effectiveness of the new filter is assessed
in a low-Earth orbit (LEO) scenario using a high-fidelity perturbation
model across a broad spectrum of orbits and conjunction geometries.
Additionally, the method is applied to space traffic management providing
a rapid and efficient means to examine the radial overlap of RSOs
in LEO and track its progression in time. 
\end{abstract}

\section*{Nomenclature}

{\renewcommand\arraystretch{1.0}
\noindent\begin{longtable*}{@{}l @{\quad=\quad} l@{}}
 $A_{v}$ & satellite cross section area, m$^{2}$\\
 $a$ & dimensionless osculating semimajor axis \\
 $a_{sp}$ & semimajor axis short-periodic component \\
 $\hat{a}$ & dimensionless mean semimajor axis \\
 $B$ & ballistic coefficient, m$^{2}\cdot$kg$^{-1}$\\
 $B^{*}$ & TLE starred ballistic coefficient, m$^{2}\cdot$kg$^{-1}$\\
 $b_{i}$ & buffer radial distance applied to the $i$-th orbit, km \\
 $C_{D}$ & drag coefficient, kg$\cdot$ m$^{-2}$ \\
 $E$ & eccentric anomaly, rad \\
 $e$ & osculating eccentricity \\
 $e_{f}$ & Cook's frozen eccentricity \\
 $e_{p}$ & proper eccentricity \\
 $e_{sp}$ & eccentricity short-periodic component \\
 $\hat{e}$ & mean eccentricity \\
 $\hat{e}_{0}$ & initial mean eccentricity \\
$h$ & altitude variation due to atmospheric drag, km\\
 $h_{0}$ & initial altitude, km\\
 $h_{min}$ & minimum altitude, km\\
 $i$ & osculating inclination, rad \\
 $i_{sp}$ & inclination short-periodic component , rad \\
 $\hat{i}$ & mean inclination, rad \\
 $J_{2}$ & second zonal spherical harmonic coefficient (oblateness)
\\
 $J_{i}$ &$i$-th zonal harmonic coefficient \\
 $k$ & angular frequency of Cook's eccentricity vector, rad$\cdot$s$^{-1}$\\
 $M$ & osculating mean anomaly, rad \\
 $M_{sp}$ & mean anomaly short-periodic component, rad \\
 $\hat{M}$ & ``mean'' mean anomaly, rad \\
 $m_{v}$ & satellite mass, kg\\
 $N$ & number of pairs\\
 $N_{FP}$ & number of false positives\\
 $N_{FN}$ & number of false negatives\\
 $N_{RP}$ & number of real (numerically obtained) positive outcomes
\\
 $N_{out}$ & number of pairs eliminated by the filter\\
 $P_{n}^{1}$ & Legendre function of order one and degree $n$ \\
 $R_{\oplus}$ & Earth radius, km \\
 $r$ & orbit radius \\
 $r_{N}$ & minimum radial distance of frozen orbit\\
 $r_{S}$ & maximum radial distance of frozen orbit \\
 $r_{f}$ & frozen orbit radius \\
 $r_{max}^{long}$ & maximum orbit radius using long-term SO theory
\\
 $r_{max}^{num}$ & maximum radius by numerical propagation\\
 $r_{max,i}$ & $i$-th orbit maximum radius reference value for
conjunction assessment , km\\
$r_{max}^{short}$ & maximum orbit radius using short-term SO theory
\\
 $r_{n}\left(n=0,1\right)$ & orbital radius at the time interval
endpoints\\
 $r_{n}^{max}\left(n=0,1\right)$ & maximum radius at the time interval
endpoints\\
 $SOR$ & space occupancy range \\
 $t$ & time, s\\
 $\beta$ & rotation angle of the proper eccentricity vector, rad\\
 $\alpha$ & rotation angle value at initial time, rad\\
 $\beta_{a}$ & constant of the exponential model fit of the atmospheric
density, km$^{-1}$\\
 $\beta_{n}\left(n=0,1\right)$ & rotation angle at time interval
endpoints, rad\\
 $\beta_{max}^{*}$ & rotation angle corresponding to the global
maximum orbit radius, rad\\
 $\beta_{Lmax}^{*}$ & rotation angle corresponding to a locally
maximum orbit radius, rad\\
 $\Delta$ & offset of the frozen orbit trajectory centroid\\
 $\varepsilon$ & short-term SO model error, km\\
 $\eta$ & filter effectiveness\\
 $(\xi,\eta)$ & mean eccentricity vector nodal components \\
 $\left(\xi_{n},\eta_{n}\right)\text{\ensuremath{\left(n=0,1\right)}}$
& mean eccentricity vector nodal components at time interval endpoints
\\
 $\theta$ & argument of latitude, rad \\
 $\hat{\theta}$ & mean argument of latitude, rad \\
 $\hat{\theta}_{max}^{*}$ & mean argument of latitude corresponding
to the global maximum orbit radius, rad\\
 $\hat{\theta}_{Lmax}^{*}$ & mean argument of latitude corresponding
to a locally maximum orbit radius, rad\\
 $\hat{\theta}_{n}^{*}\left(n=0,1\right)$ & mean argument of latitude
of the maximum radius at time endpoints, rad\\
 $\mu$ & Earth gravitational parameter, km$^{3}\cdot$s$^{-2}$\\
 $\nu$ & true anomaly, rad \\
 $\rho$ & atmospheric density, kg$\cdot$m$^{-3}$\\
 $\rho_{FN}$ & false negatives to detected real positives ratio\\
 $\rho_{FP}$ & false positives to detected real positives ratio\\
 $\bar{\rho}$ & constant of the exponential model fit of the atmospheric
density, kg$\cdot$m$^{-3}$\\
 $\tau$ & dimensionless time \\
 $\Omega$ & osculating right ascension of the ascending node, rad
\\
 $\Omega_{sp}$ & right ascension of the ascending node short-periodic
component, rad \\
 $\hat{\Omega}$ & mean right ascension of the ascending node, rad
\\
 $\omega$ & osculating argument of periapsis, rad \\
 $\omega_{sp}$ & argument of periapsis short-periodic component
, rad \\
 $\hat{\omega}$ & mean argument of periapsis, rad \\
 $\hat{\omega}_{0}$ & initial mean argument of periapsis, rad \\
 \end{longtable*}}

\section{Introduction\label{sec:intro}}

\lettrine{T}{he} space environment is becoming ever more crowded,
both because of the deployment of large constellations of small to
medium-sized satellites by several companies, as well as the growing
number of space debris. Over the next decade, more than 20,000 satellites
are projected to be placed in orbit, as proposed by approximately
two dozen companies. Providing perspective for the magnitude of this
number, it is worth noting that since the inception of the space age,
fewer than 8,100 payloads have been launched into Earth orbit \cite{STM}.
On the other hand, the quantity of small, undetectable orbital debris
is increasing progressively and exceeds the number of large, cataloged
space objects by several orders of magnitude \cite{small_debris},
further compounding the risk of collisions. Consequently, the probability
of a fatal incident resulting in satellite loss is increasing and
not negligible \cite{sgobba} and if such an event were to occur,
it would not only destroy the two objects involved but also generate
a significant amount of debris, which can in turn produce more collisions.
The impact between an Iridium satellite and COSMOS 2251, which took
place in 2009 is a clear example of such circumstances \cite{IC}.
To mitigate the risk of collisions, the initial step involves detecting
close encounters between satellites of interest and other objects.
A close encounter, or critical conjunction, occurs when the expected
orbits of two satellites result in a minimum approach distance below
a few kilometers. Once a critical conjunction is identified, the threat
level is assessed by computing the probability of collision. Finally,
if this probability is above a given threshold, a collision avoidance
maneuver is carried out \cite{Krage}.

Considering the substantial number of Resident Space Objects (RSOs),
this process is computationally intensive. The ``all on all'' conjunction
screening problem is seldom addressed operationally due to its inherent
difficulties; comparing an entire catalog of more than 20,000 objects
against itself leads to screening over 400 million pairs. This challenge
is poised to intensify in the future due to increased space traffic
and enhanced observation capabilities, which will significantly expand
the catalog of objects and, consequently, the number of potential
conjunction pairs \cite{Stevenson2023}. Nonetheless, prior research
has explored various strategies to mitigate this calculation load,
such as filtering processes and parallelization \cite{Escobar,Hall,Healy,Wood}.
Specifically, applying filters to exclude RSO pairs with negligible
collision risk emerges as one of the most widespread computational
acceleration techniques for conjunction analyses \cite{kerr2021state}.
Literature proposes several filter types, notably apogee-perigee (AP),
orbit path, and time filters. The first two are purely based on the
geometry of the orbit pair \cite{CF} while the latter takes phasing
into account. The AP filter eliminates pairs with non-overlapping
altitude bands, precluding possible collisions. The orbit path filter
excludes pairings with a consistently greater distance than the safety
threshold. In contrast, the time filter relies on temporal coincidence
of orbits which are close. The fact that two orbits are within a threshold
distance of each other does not guarantee a conjunction; the satellites
traversing these orbits must also be within the specified proximity
simultaneously \cite{FILT}.

The AP filter stands out for its straightforwardness among the trio
of mentioned screening methods. It selectively excludes pairs of objects
whose Earth-centric radii cannot intersect, determined by the minimal
and maximal radial values derived from perigee and apogee calculations,
respectively. Crucially, this evaluation necessitates only a single
computation per object, specifically using the explicit formulas to
determine perigee and apogee radii, which significantly reduces the
processing demand. This adjustment allows the ``all on all' conjunction
analysis to scale linearly with the size of the population, rather
than quadratically. Nevertheless, the reliance of this method on a
Keplerian framework introduces a notable limitation. Perturbations,
chiefly among them gravitational harmonics, lead to variations in
orbital elements, subsequently causing alterations in perigee and
apogee distances. Thus, introducing an adjustable buffer zone is essential
to accommodate these changes. Establishing such a safety threshold,
however, is inherently heuristic and demands careful consideration
\cite{BUF}. An undersized buffer may overlook actual conjunction
threats, whereas an oversized one could unnecessarily retain numerous
non-critical pairings. This situation underscores the screening methods'
vulnerability to two types of errors: false positives, where non-threatening
pairs are marked for further evaluation, and false negatives, where
potential conjunctions are prematurely dismissed.

A few studies aim to refine the classical AP filter. For example,
Casanova's Filter I \cite{Casanova} addresses two problems associated
with Hoot's AP version \cite{CF}. Initially, it computes the apogee
and perigee for each object using osculating elements, thus addressing
the first issue of sensitivity to both short-term and long-term perturbations.
The second concern is the risk of false positives during the comparison
of apogee and perigee for two objects at a specified time, especially
if one is an active satellite executing a maneuver. This is mitigated
by utilizing ephemerides tables to determine equinoctial orbital elements,
which are then linearly interpolated to estimate these parameters
at any given moment. The core strategy of this filter involves shifting
from direct apogee and perigee calculations to determining the geocentric
global extrema of the distance at a precise time. Woodburn \cite{Wood}
suggests two additional enhancements to this approach. The initial
strategy adds padding to the detection threshold and examines each
trajectory at the start and end of the analysis interval, assessing
the radial distance variations for each object. This dual-point sampling
aims to account for apogee or perigee trends. The alternate strategy,
akin to Filter I, opts for selective sampling across all trajectories
to ascertain the minimum and maximum radial distances of each one,
albeit at a higher computational cost.

A recent study by Bombardelli et al. \cite{SO} introduces the concept
of space occupancy (SO) as a means to organize the distribution of
Resident Space Objects (RSOs) into non-overlapping orbital shells,
thereby minimizing the number of required collision avoidance maneuvers.
The work was focused on exploring potential solutions for optimizing
RSO distribution over the long term (a few months or longer), considering
SO on a time scale sufficient for most LEO objects to complete a full
rotation of the line of apsides. That concept, denoted as "long-term
SO" in the present work, was shown to be closely linked to the geometry
of frozen orbits and the notion of proper eccentricity, first introduced
by Cook in 1966 \cite{cook1966perturbations}.

For conjunction screening, which operates within significantly shorter
time horizons (days) compared to the precession period of the line
of apsides (typically, months), the established SO theory, as detailed
in \cite{SO}, proves to be overly conservative for most scenarios.
By addressing this gap, the central contribution of the current work
is the introduction and development of a ``short-term SO'' theory,
which offers a tailored methodology for accurately and swiftly estimating
the altitude range within which a satellite operates over specific,
short timeframes, achieved through solving a quartic equation. This
refined approach supports the creation of a novel conjunction filter,
termed the Space Occupancy conjunction filter (SO-filter), which surpasses
the traditional AP one by precisely accounting for the influence of
zonal harmonics on a trajectory of the RSO. The SO-filter, based on
a zonal-perturbed two-body problem, emerges as a superior alternative
for conjunction analysis. Additionally, this work applies SO principles
to the broader context of space traffic management, culminating with
a succinct review of the evolution of space population, presented
at the end of the study.

The structure of the article is as follows. First, Section~\ref{sec:sot}
revisits the SO theory and lays the theoretical groundwork for short-term
SO, essential for the operational advancements introduced later. The
results are validated using a high-fidelity propagator. Next, Section~\ref{sec:filter}
describes the development and analytical implementation of several
filters based on the short-term SO model. Their performance is then
compared to that of the classical AP filter, focusing on reliability
and effectiveness. Finally, Section~\ref{sec:evolution} offers a
concise analysis of the evolution of the space population from 2005
to the present, employing the developed SO tools. The article concludes
with some final remarks in Section~\ref{sec:conclusions}.

\section{Space Occupancy Theory\label{sec:sot}}

This section starts by revisiting the foundational concepts of Space
Occupancy (Section~\ref{sec:long}), referred to here as long-term
SO due to its focus on long time intervals. In contrast, the primary
contribution of this work is centered on short-term SO, which is more
useful for operational tasks such as conjunction analysis. The novel
short-term SO theory is introduced in Section~\ref{sec:STSO}, which
includes a comprehensive numerical validation using a high-fidelity
propagator.

\subsection{Long-term Space Occupancy}

\label{sec:long}

SO was introduced in Bombardelli et al. \cite{SO} as the complete
geometric domain that a satellite occupies as it traverses its nominal
orbit, under the influence of environmental perturbations, during
a defined interval. This concept parallels the reachability domain
\cite{Holzinger2009}, used in operations such as spacecraft rendezvous
or formation flying \cite{sanchez2021event}, where thrust accelerations
define accessible spatial regions. Yet, in contrast, SO is uniquely
governed by environmental perturbations. Notably, when the interval
under study coincides with or exceeds the $J_{2}$-induced apsides
precession period, SO becomes intimately connected with the geometry
of frozen orbits and the notion of proper eccentricity, concepts originally
introduced by Cook \cite{cook1966perturbations}. These elements provide
analytical insights into RSO orbital dynamics, enhancing our comprehension
of SO. The contributions of Bombardelli et al. \cite{SO} and some
notions from the work of Cook \cite{cook1966perturbations} on frozen
orbit theory are succinctly reviewed below for the reader's convenience.

Consider the Earth's radius, $R_{\oplus}$, as the reference length
unit and $1/n_{\oplus}$ as the reference time, with $n_{\oplus}$
the mean motion of a circular orbit of radius $R_{\oplus}$. Denote
by $\tau$ the dimensionless time thus defined. Let $\hat{e}$, $\hat{\omega}$,
$\hat{a}$, and $\hat{i}$ represent the mean values---averaged over
the mean anomaly---of the eccentricity, argument of periapsis, semimajor
axis, and inclination, respectively \cite{vallado2001fundamentals}.

The foundational work of Cook \cite{cook1966perturbations} describes
the behavior of the mean eccentricity vector of an orbit influenced
by $J_{2}$ and an arbitrary sequence of odd zonal harmonics, with
components $(\xi,\eta)$ in the nodal frame given by 
\begin{equation}
\xi(\tau)=\hat{e}(\tau)\cos\hat{\omega}(\tau),\qquad\eta(\tau)=\hat{e}(\tau)\sin\hat{\omega}(\tau).\label{eq:eccentricity_vector}
\end{equation}

\begin{figure}[t!]
\centering \includegraphics[width=0.5\textwidth]{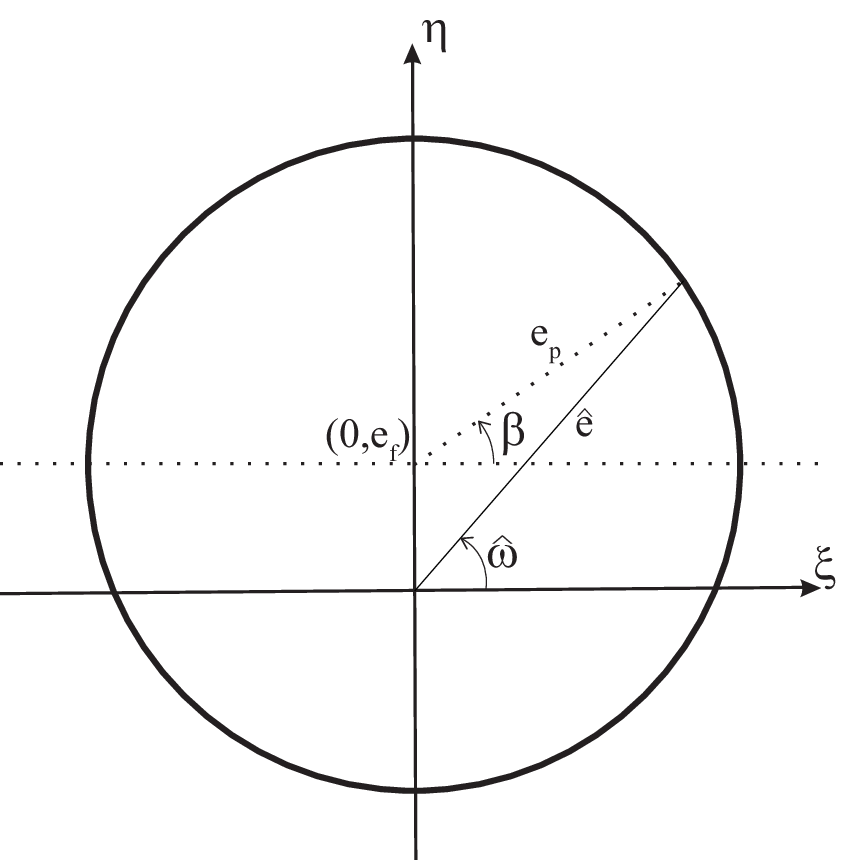}
\caption{Motion of the eccentricity vector in the ($\xi,\eta$) plane.}
\label{Cook} 
\end{figure}

For orbits of small eccentricity, these vectors trace a circular path
(``Cook's circle'') defined by 
\begin{eqnarray}
\xi(\tau) & = & e_{p}\cos\beta\left(\tau\right),\label{eq:sol_Cook_xi}\\
\eta(\tau) & = & e_{p}\sin\beta\left(\tau\right)+e_{f},\label{eq:sol_Cook_eta}
\end{eqnarray}
where $e_{f}$, $e_{p}$, and $\beta\left(\tau\right)$ symbolize
the frozen eccentricity, proper eccentricity, and rotation angle,
respectively, as illustrated in Fig. \ref{Cook}. Notably, $(0,e_{f})$
represents the center of Cook's circle with $e_{p}$ as its radius,
indicating that $e_{f}$ can be either positive or negative, while
$e_{p}$ remains nonnegative. Cook's theory meticulously provides
the analytical expressions for these parameters \cite{cook1966perturbations,SO}:
\begin{eqnarray}
e_{f} & = & k^{-1}\hat{a}^{-3/2}\sum_{n=1}^{N}\frac{J_{2n+1}}{\hat{a}_{0}^{2n+1}}\frac{n}{\left(2n+1\right)\left(n+1\right)}P_{2n+1}^{1}(0)P_{2n+1}^{1}(\cos\hat{i}),\label{eq:e_f}\\
e_{p} & = & \sqrt{\left(\hat{e}_{0}\sin\hat{\omega}_{0}-e_{f}\right)^{2}+\hat{e}_{0}^{2}\cos^{2}\hat{\omega}_{0}},\label{ep}\\
\sin\alpha & = & \frac{\hat{e}_{0}\sin\hat{\omega}_{0}-e_{f}}{e_{p}},\label{eq:sin_a}\\
\cos\alpha & = & \frac{\hat{e}_{0}\cos\hat{\omega}_{0}}{e_{p}},\label{eq:cos_a}\\
\beta\left(\tau\right) & = & k\tau+\alpha,\\
k & = & \frac{3J_{2}}{\hat{a}^{7/2}}\left(1-\frac{5}{4}\sin^{2}\hat{i}\right),\label{eq:k_def}
\end{eqnarray}
where $P_{n}^{1}$ denotes the associated Legendre function of the
first order and degree $n$, and $J_{2n+1}$ are the odd zonal harmonic
coefficients, see e.g. \cite{vallado2001fundamentals}. The term $k$,
as defined in Eq.~\eqref{eq:k_def}, modulates the rotation of the
eccentricity vector. Initial conditions $\hat{e}_{0}$ and $\hat{\omega}_{0}$
represent the mean eccentricity and argument of periapsis at $\tau=0$.

From these expressions, the mean eccentricity and argument of periapsis
can be readily computed as: 
\begin{eqnarray}
\hat{e}\left(\tau\right) & = & \sqrt{e_{f}^{2}+e_{p}^{2}+2e_{f}e_{p}\sin\beta\left(\tau\right)},\label{eq:e_m1}\\
\hat{\omega}\left(\tau\right) & = & \mathrm{atan2}\left(\eta\left(\tau\right),\xi\left(\tau\right)\right).\label{eq:e_m2}
\end{eqnarray}

Cook's theory lays the groundwork for the comprehension of frozen
orbits, which result from a combination of orbital elements such that
the proper eccentricity $e_{p}$ in Eq. (\ref{eq:e_f}) becomes zero,
thus making the mean eccentricity of the orbit constant and equal
to $\vert e_{f}\vert$. A significant contribution of SO theory is
its application of frozen orbits as a framework to ascertain the long-term
radial behavior of generic orbits with small eccentricities. This
involves first determining the maximum and minimum radii of frozen
orbits. Considering that the values of $\hat{a}$ and $\hat{i}$ remain
constant, a polar equation for the radius of a frozen orbit as a function
of the mean argument of latitude $\hat{\theta}$, expanded to first-order
eccentricity, is derived in \cite{SO}: 
\begin{equation}
r_{f}\left(\hat{\theta}\right)=\hat{a}\left(1-e_{f}\sin\hat{\theta}\right)+\frac{J_{2}}{4\hat{a}}\left[\left(9+\cos2\hat{\theta}\right)\sin^{2}\hat{i}-6\right].\label{eq:polar_frozen}
\end{equation}

The maximum and minimum values of the radius\footnote{It is crucial to differentiate between maximum and minimum radius
of the SO and apogee or perigee concepts. The former are global, potentially
conservative bounds for occupied orbital space, whereas the latter,
can refer to mean or osculating values. Osculating values change continuously,
and mean values may not provide precise bounds, as demonstrated in
subsequent sections.} as given by Eq. (\ref{eq:polar_frozen}) , denoted as $r_{N}$ (minimum
radius) and $r_{S}$ (maximum radius), are described by\footnote{In \cite{SO}, the consideration of harmonics up to order 3 in the
frozen eccentricity computation results in $e_{f}$ always being positive,
leading to the exclusion of cases near the critical inclination, where
$e_{f}$ could be negative. To address these cases in Eq. (\ref{eq:r_N})
and (\ref{eq:r_S}), the absolute value has been incorporated.}: 
\begin{eqnarray}
r_{N} & \simeq & \hat{a}\left(1-|e_{f}|\right)+\dfrac{J_{2}\left(4\sin^{2}\hat{i}-3\right)}{2\hat{a}},\label{eq:r_N}\\
r_{S} & \simeq & \hat{a}\left(1+|e_{f}|\right)+\dfrac{J_{2}\left(4\sin^{2}\hat{i}-3\right)}{2\hat{a}}.\label{eq:r_S}
\end{eqnarray}

\begin{figure}[t!]
\centering \includegraphics[width=0.7\textwidth]{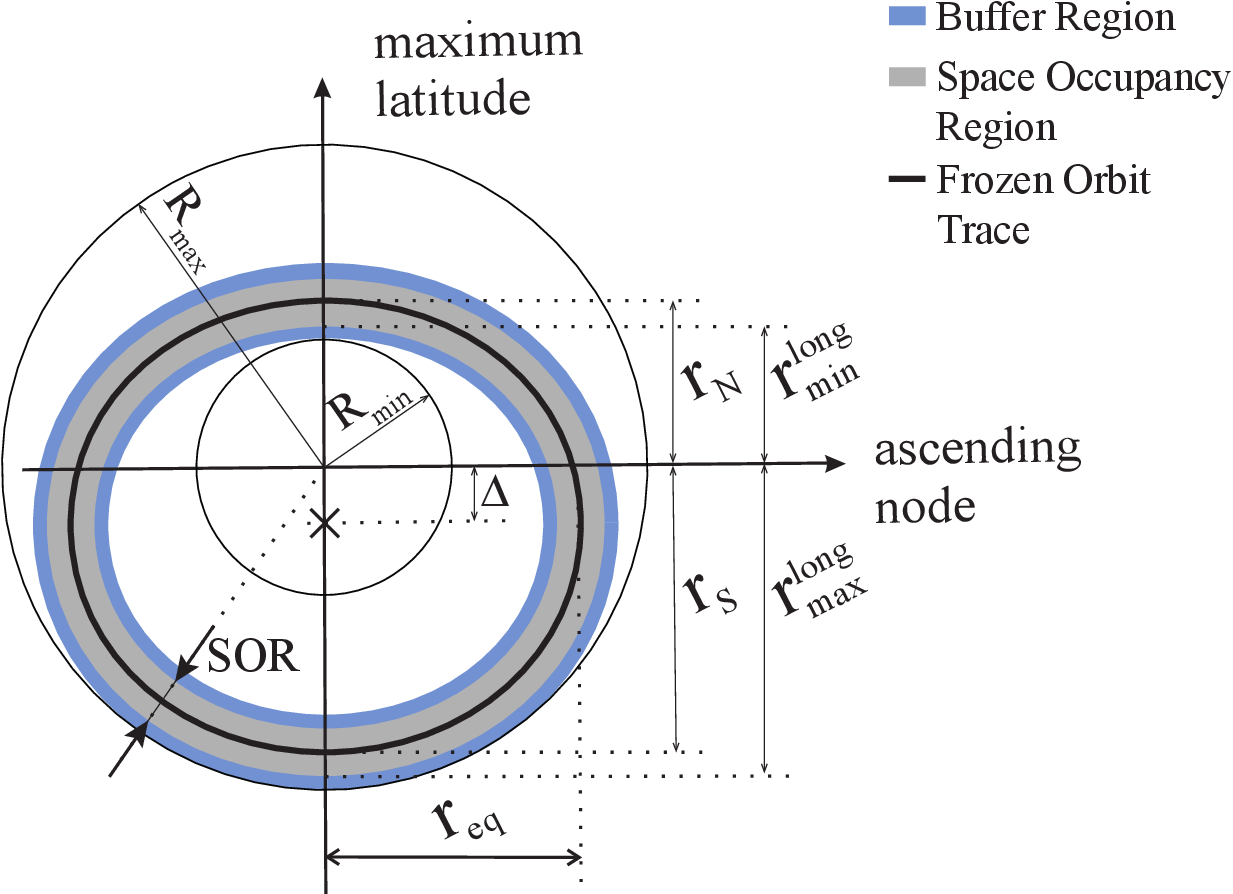}
\caption{\label{fig:fig1-1}Space Occupancy Geometry.}
\end{figure}

Ordinarily, the minimum radius is attained at the orbital ``North''
and the maximum at ``South'' for Earth orbits, due to the specific
values of Earth's $J_{2}$ and odd zonal harmonics making $e_{f}$
in Eq. (\ref{eq:e_f}) positive\footnote{Here,``North'' and ``South'' denote the northernmost and southernmost
positions of the orbit, respectively, achieved when the argument of
latitude reaches $\pi/2$ or 90 degrees (respectively $3\pi/2$ or
270 degrees)}. These extrema can switch places if $e_{f}$ is negative, a scenario
that can only arise (for Earth) in orbits nearing the critical inclination
(63.4 or 116.6 deg). Note that in those cases the offset $\Delta$,
defined by Eq. (\ref{eq:delta}) and depicted in Fig. \ref{fig:fig1-1},
which is typically negative (southward) can become zero or positive
(northward). 
\begin{eqnarray}
\Delta & = & r_{N}-r_{S}=-2\hat{a}e_{f}.\label{eq:delta}
\end{eqnarray}

Utilizing frozen orbits, SO theory demonstrated that the radial bounds
for the long-term SO of \emph{any} small eccentricity orbit influenced
by $J_{2}$ and arbitrarily many odd zonal harmonics can be calculated
as follows based on the corresponding frozen orbit with the same values
of $\hat{a}$ and $\hat{i}$, incorporating an additional quantity,
the \textit{space occupancy radius} (SOR): 
\begin{eqnarray}
r_{min}^{long} & = & r_{N}-\frac{SOR}{2}=r_{N}-\hat{a}e_{p},\label{eq:r_min_LONG}\\
r_{max}^{long} & = & r_{S}+\frac{SOR}{2}=r_{S}+\hat{a}e_{p},\label{eq:r_max_LONG}
\end{eqnarray}
where $r_{min}^{long}$ and $r_{max}^{long}$ represent, respectively,
the long-term minimum and maximum radial bounds within which the space
occupancy of an orbit is confined over an extended period. The SOR,
as illustrated in Fig.~\ref{fig:fig1-1}, quantifies the range between
the maximum and minimum orbital radii, which are centered around the
corresponding frozen orbit, and is expressed as $SOR\simeq2\hat{a}e_{p}$.


\begin{figure}[hbt!]
\includegraphics[width=1\textwidth]{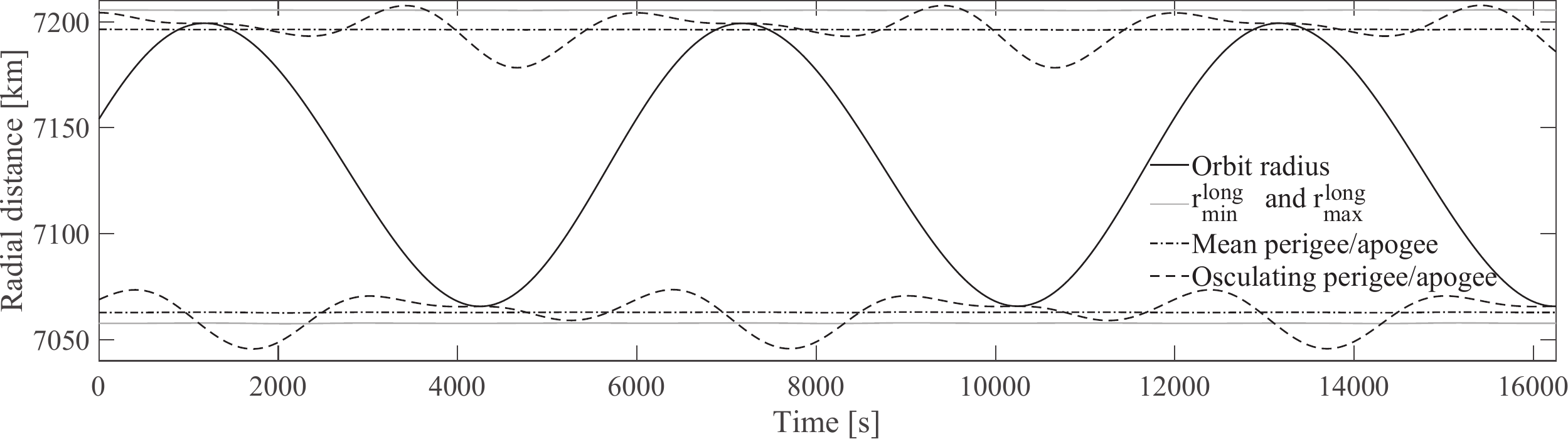} \caption{Comparison of mean and osculating apogee/perigee against long-term
SO bounds with actual radius variation for a LEO orbit.}
\label{radii} 
\end{figure}

To illustrate the SO radial bounds, Fig.~\ref{radii} compares the
calculated $r_{max}^{long}$ and $r_{mmin}^{long}$ values with the
mean apogee and perigee for a typical LEO orbit. It also displays
the actual radius, computed with a propagator, alongside the osculating
perigee and apogee changes over time. This comparison highlights that
the orbital radius consistently remains within the bounds predicted
by SO theory, unlike the mean apogee/perigee, which may not provide
precise bounds. The depicted orbit is characterized by its initial
osculating elements: $a=7,136.6~\mathrm{km},e=0.0095,i=1.2723~\mathrm{rad},\omega=1.004~\mathrm{rad},\Omega=2.0246~\mathrm{rad},M=1.8230~\mathrm{rad}$.
This figure is generated using the high-fidelity propagator described
in Section~\ref{sec:validation}.

\subsection{Short-term Space Occupancy}

\label{sec:STSO}

When the space occupancy radius needs to be evaluated for time intervals
shorter than the line of apsides precession period, as for instance
would be the case for conjunction analysis, the long-term SO bounds
determined by Eqs.~\eqref{eq:r_min_LONG} and \eqref{eq:r_max_LONG}
may prove overly conservative. This motivates a refined space occupancy
framework to accurately capture the evolution of the radial bounds
of a generic orbit over such shorter timeframes. For this purpose,
a polar equation for non-frozen orbits is obtained as a means to provide
lower and upper radial bounds over a selected time interval by solving
a simple quartic equation. The analytically computed bounds are validated
in Section~\ref{sec:validation} using a high-fidelity propagator.

\subsubsection{A Polar Equation for Non-frozen Orbits}

\label{sec:time}

The approximate evolution of the radius for a non-frozen orbit with
small eccentricity, using a first-order eccentricity expansion, is
derived from reference \cite{SO}, 
\begin{equation}
r\approx\hat{a}\left(1-\hat{e}\cos(\hat{\theta}-\hat{\omega})\right)+\frac{J_{2}}{4\hat{a}}\left[\left(9+\cos(2\hat{\theta})\right)\sin^{2}\hat{i}-6\right].\label{eqn-hatr}
\end{equation}

In order to determine all local maxima and minima of (\ref{eqn-hatr}),
its time derivative is computed and set to zero: 
\begin{equation}
\frac{dr\left(\tau\right)}{d\tau}=\frac{\partial r}{\partial\hat{e}}\frac{d\hat{e}}{d\tau}+\frac{\partial r}{\partial\hat{\omega}}\frac{d\hat{\omega}}{d\tau}+\frac{\partial r}{\partial\hat{M}}\frac{d\hat{M}}{d\tau}=0,
\end{equation}
leading to: 
\begin{equation}
-\hat{a}\frac{d\hat{e}}{d\tau}\cos\hat{M}+\hat{a}\hat{e}\sin\hat{M}\frac{d\hat{M}}{d\tau}-\frac{J_{2}\sin^{2}\hat{i}}{2\hat{a}}\sin\left(2\hat{M}+2\hat{\omega}\right)\left(\frac{d\hat{M}}{d\tau}+\frac{d\hat{\omega}}{d\tau}\right)=0.\label{partial_M}
\end{equation}

By differentiating Eqs.~(\ref{eq:sol_Cook_xi})--(\ref{eq:sol_Cook_eta})
and (\ref{eq:e_m1})--(\ref{eq:e_m2}), the time derivatives of $\hat{\omega}$
and $\hat{e}$ are: 
\begin{eqnarray}
\frac{d\hat{\omega}}{d\tau} & \simeq & \frac{1}{\hat{e}^{2}}\left(\frac{d\eta}{d\tau}\xi-\eta\frac{d\xi}{d\tau}\right),\label{domega}\\
\frac{d\xi}{d\tau} & = & -e_{p}k\sin\beta,\label{dxi}\\
\frac{d\eta}{d\tau} & = & e_{p}k\cos\beta,\label{deta}\\
\frac{d\hat{e}}{d\tau} & \simeq & \frac{1}{2\hat{e}}\left(2\xi\frac{d\xi}{d\tau}+2\eta\frac{d\eta}{d\tau}\right),\label{de}
\end{eqnarray}
and the time derivative of the mean anomaly, retaining only secular
terms and neglecting the long-periodic contribution, is \cite{Kozai}:
\begin{equation}
\frac{d\hat{M}}{d\tau}\simeq\sqrt{\frac{1}{\hat{a}^{3}}}+\frac{3J_{2}}{2\hat{a}^{7/2}(1-\hat{e}^{2})^{3/2}}\left(1-\frac{3}{2}\sin^{2}\hat{i}\right).\label{dM}
\end{equation}

Solving Eq. \eqref{partial_M} with Eqs. (\ref{domega}--\ref{dM})
provides a highly accurate yet computationally intensive approach
for determining short-term space occupancy bounds.

\subsubsection{Short-term SO Quartic Equation}

\label{sec:quartic}

A more efficient method to determine short-term SO bounds can be devised
by reformulating Eq.(\ref{eqn-hatr}). Subsequently, expanding $\cos(\hat{\theta}-\hat{\omega})=\cos\hat{\theta}\cos\hat{\omega}+\sin\hat{\theta}\sin\hat{\omega}$
and using Eq. (\ref{eq:eccentricity_vector}), the first term of Eq.
(\ref{eqn-hatr}) can be written as: 
\begin{eqnarray}
\hat{a}\left(1-\hat{e}\cos(\hat{\theta}-\hat{\omega})\right) & = & \hat{a}\left(1-\left(\xi\cos\hat{\theta}+\eta\sin\hat{\theta}\right)\right)\nonumber \\
 & = & \hat{a}\left(1-e_{p}\left(\cos\beta\cos\hat{\theta}+\sin\beta\sin\hat{\theta}\right)-e_{f}\sin\hat{\theta}\right)\nonumber \\
 & = & \hat{a}\left(1-e_{p}\cos(\hat{\theta}-\beta)-e_{f}\sin\hat{\theta}\right),\label{eqn-hatahate4}
\end{eqnarray}
where Eqs. (\ref{eq:sol_Cook_xi})--(\ref{eq:sol_Cook_eta}) have
been employed. Incorporating Eq.~\eqref{eqn-hatahate4} into Eq.~\eqref{eqn-hatr}
yields an alternative expression for $r$, directly linking it to
$\hat{\theta}$ and $\beta$: 
\begin{equation}
r\left(\hat{\theta},\beta\right)\approx\hat{a}\left(1-e_{p}\cos(\hat{\theta}-\beta)-e_{f}\sin\hat{\theta}\right)+\frac{J_{2}}{4\hat{a}}\left[\left(9+\cos(2\hat{\theta})\right)\sin^{2}\hat{i}-6\right].\label{eqn-hatr2}
\end{equation}

Given the slow, long-term movement of $\beta=k\tau+\alpha$ and the
more rapid changes in the argument of latitude $\hat{\theta}$, it
becomes evident that $\hat{\theta}$ crosses its entire range $[0,2\pi)$
significantly faster than $\beta$. This observation justifies treating
$\beta$ and $\hat{\theta}$ as independent variables for the purposes
of short-term space occupancy thereby converting the time-domain analysis
in Section~\ref{sec:time} into a simpler multivariable optimization
problem. As $r$ represents a continuous, differentiable function
in both $\beta$ and $\hat{\theta}$, basic calculus principles dictate
that its extremal values must occur at either critical (zero-derivative)
points or the domain boundaries.

The following is only focused in the problem of finding the maximum
of $r$, noting that a similar procedure should be followed to address
finding the minimum of $r$.

In the long-term SO problem, both $\hat{\theta}$ and $\beta$ are
periodic and therefore the global maximum is always found at critical
points as:

\[
r_{max}^{long}=r\left(\hat{\theta}_{max}^{*},\beta_{max}^{*}\right).
\]

The computation of the critical point giving rise to the global maximum
$\left(\hat{\theta}_{max}^{*},\beta_{max}^{*}\right)$, as well as
other critical point that correspond to relative maxima $\left(\hat{\theta}_{Lmax}^{*},\beta_{Lmax}^{*}\right)$,
as a function of the mean orbital elements is deferred to Appendix
A for clarity and conciseness.

Conversely, when addressing the short-term SO problem, and denoting
with $\tau_{f}$ the final time, the variable $\hat{\theta}$ crosses
its full range $[0,2\pi)$ while $\beta$ , illustrated in Fig. \ref{Cook},
spans a shorter arc $\beta\in[\beta_{0},\beta_{1}]$ with:

\[
\beta_{0}=\alpha,\qquad\beta_{1}=\alpha+k\tau_{f},
\]
where $\alpha$ is defined by Eqs. \eqref{eq:sin_a} and \eqref{eq:cos_a}.

Three possible scenarios arise. If the interval $[\beta_{0},\beta_{1}]$
contains the global maximum value $\beta_{max}^{*}$ then the solutions
for the short-term and long-term SO problem coincide. If the interval
contains a local maximum value $\beta_{Lmax}^{*}$ the radius becomes
maximum either at this value or at one of the interval endpoints.
Otherwise, the maximum corresponds to one of the interval endpoints.
Therefore, the short-term SO maximum can be computed as: 
\begin{flushleft}
\begin{eqnarray}
r_{max}^{short} & \left\{ \begin{array}{lll}
r_{max}^{long} & \mathrm{if} & \beta_{max}^{*}\in[\beta_{0},\beta_{1}]\\
\max\left\{ r(\beta_{Lmax}^{*},\hat{\theta}_{Lmax}^{*}),r_{0}^{max},r_{1}^{max}\right\}  & \mathrm{if} & \beta_{Lmax}^{*}\in[\beta_{0},\beta_{1}]\\
\max\left\{ r_{0}^{max},r_{1}^{max}\right\}  & \mathrm{otherwise}
\end{array}\mathrm{\land}\;\beta_{max}^{*}\notin[\beta_{0},\beta_{1}]\right.\label{eq:r_max_short}
\end{eqnarray}
where:
\par\end{flushleft}

\[
r_{n}^{max}=\max\left\{ r_{n}(\hat{\theta}),\hat{\theta}=0..2\pi\right\} \qquad n=0,1,
\]
with

\begin{equation}
r_{n}(\hat{\theta})=r_{n}(\hat{\theta},\beta_{n})\approx\hat{a}\left(1-e_{p}\cos(\hat{\theta}-\beta_{n})-e_{f}\sin\hat{\theta}\right)+\frac{J_{2}}{4\hat{a}}\left[\left(9+\cos(2\hat{\theta})\right)\sin^{2}\hat{i}-6\right].\label{eq:r_n}
\end{equation}

The value of $\hat{\theta}$, denoted with $\hat{\theta}_{n}^{*}$
, that maximizes $r_{n}(\hat{\theta})$, can be obtained by computing
the derivative of Eq. \eqref{eq:r_n} and setting it to zero. In this
fashion:

\[
r_{n}^{max}=r_{n}(\hat{\theta}_{n}^{*}),
\]
where:

\begin{equation}
0=\frac{dr_{n}}{d\hat{\theta}}\left(\hat{\theta}_{n}^{*}\right)=\hat{a}\left(e_{p}\sin\left(\hat{\theta}_{n}^{*}-\beta_{n}\right)-e_{f}\cos\hat{\theta}_{n}^{*}\right)-\frac{J_{2}k}{2\hat{a}}\sin\left(2\hat{\theta}_{n}^{*}\right).\label{eqn-sa}
\end{equation}

Significantly, Eq.~\eqref{eqn-sa} can be reduced to a polynomial.
Employing the classical substitution $x_{n}=\tan(\hat{\theta}_{n}^{*}/2)$,
which establishes a bijective relationship between $\mathbb{R}$ and
$(-\pi,\pi)$, leads to: 
\begin{equation}
0=\hat{a}\left(e_{p}\frac{2x_{n}}{1+x_{n}^{2}}\cos(\beta_{i})-e_{p}\frac{1-x_{n}^{2}}{1+x_{n}^{2}}\sin(\beta_{i})-e_{f}\frac{1-x_{n}^{2}}{1+x_{n}^{2}}\right)-\frac{J_{2}k}{\hat{a}}\frac{2x_{n}(1-x_{n}^{2})}{(1+x_{n}^{2})^{2}}.\label{eqn-sax}
\end{equation}
Cross-multiplying Eq.~\eqref{eqn-sax} by $(1+x^{2})^{2}$ results
in the quartic polynomial: 
\begin{equation}
0=-\left(\hat{a}e_{p}\sin\beta_{n}+\hat{a}e_{f}\right)+2x_{n}\left(\hat{a}e_{p}\cos\beta_{n}-\frac{J_{2}\sin^{2}\hat{i}}{\hat{a}}\right)+2x_{n}^{3}\left(\hat{a}e_{p}\cos\beta_{n}+\frac{J_{2}\sin^{2}\hat{i}}{\hat{a}}\right)+x_{n}^{4}\left(\hat{a}e_{p}\sin\beta_{n}+\hat{a}e_{f}\right),\label{eqn-sax2}
\end{equation}
or, in compact form: 
\begin{equation}
x_{n}^{4}+Px_{n}^{3}+Qx_{n}-1=0,\label{eq:SOquarticequation}
\end{equation}
with:

\[
P=\frac{2\hat{a}^{2}\hat{e}_{n}\cos\hat{\omega}_{n}+2J_{2}\sin^{2}\hat{i}}{\hat{a}^{2}\hat{e}_{n}\sin\hat{\omega}_{n}},\quad Q=\frac{2\hat{a}^{2}\hat{e}_{n}\cos\hat{\omega}_{n}-2J_{2}\sin^{2}\hat{i}}{\hat{a}^{2}\hat{e}_{n}\sin\hat{\omega}_{n}},
\]
where the following identities have been employed:

\[
e_{p}\cos\beta_{n}=\xi_{n}=\hat{e}_{n}\cos\hat{\omega}_{n}=\hat{e}\left(\beta_{n}\right)\cos\hat{\omega}\left(\beta_{n}\right),
\]
\[
e_{p}\sin\beta_{n}+e_{f}=\eta_{n}=\hat{e}\left(\beta_{n}\right)\sin\hat{\omega}\left(\beta_{n}\right).
\]

Eq.\eqref{eq:SOquarticequation} can have up to four real solutions,
and can be solved analytically, via Ferrari's method, or through efficient
numerical strategies, such as Newton's or Halley's methods.

The consideration of a potential root at infinity requires examining
the limit of the polynomial in Eq.~\eqref{eqn-sax} when the denominator
of both $P$ and $Q$ becomes zero. This scenario arises exclusively
if $\sin\hat{\omega}_{i}=0$, implying solutions at $x=0$ and $x=\infty$,
corresponding to $\hat{\theta}_{n}^{*}=0$ and $\hat{\theta}_{n}^{*}=\pi$,
respectively.


Finally, both the radius and the second derivative (used for discerning
the nature of the critical points) can be expressed directly in terms
of $x$ as: 
\begin{eqnarray}
r_{n} & = & \hat{a}\left(1-\hat{e}_{n}\frac{(1-x^{2})\cos\hat{\omega}_{n}-(2x)\sin\hat{\omega}_{n}}{1+x^{2}}\right)+\frac{J_{2}}{4\hat{a}}\left[\left(9+\frac{1-6x^{2}+x^{4}}{(1+x^{2})^{2}}\right)k-6\right],\label{eq:r_x}\\
\frac{\partial^{2}r_{n}}{\partial\hat{\theta}^{2}} & = & \hat{a}\hat{e}_{n}\frac{(1-x^{2})\cos\hat{\omega}_{n}-(2x)\sin\hat{\omega}_{n}}{1+x^{2}}-\frac{J_{2}k}{\hat{a}}\frac{1-6x^{2}+x^{4}}{(1+x^{2})^{2}}.
\end{eqnarray}

This approach significantly enhances the efficiency of the short-term
SO bounds computation.

\subsubsection{Short-term SO Validation}

\label{sec:validation}

The analytical short-term SO model developed in Section~\ref{sec:STSO}
has been tested and validated against a high-fidelity numerical propagation
model. The latter includes a 23$\times$23 geopotential model\footnote{The selection of order 23 was the result of extensive numerical testing.
Such order balances accuracy with computational efficiency, and higher
orders do not markedly change the results \cite{Rivero2022FastOrbit}.} along with lunisolar third-body perturbations, and accounts for Earth's
geoid precession and nutation. The Earth orientation, the values of
the harmonic coefficients, as well as the position of the Sun and
Moon are obtained from the corresponding SPICE kernels. Note that
the analytical short-term SO model is constructed on a zonal problem
with the earth polar axis aligned with the z axis of the J2000 inertial
frame.

Concerning the applicability limits of the proposed theory, it is
important to underline that orbits with eccentricities higher than
0.1 and apogee radii exceeding 40,000 km have been left out. Including
these orbits would violate the simplifying assumptions at the basis
of Cook's theory, which is a first order theory in eccentricity and
neglects third body perturbations, and may result in unacceptable
errors. When analyzing the space occupancy of these more eccentric
and/or near-GEO orbits a more refined numerical analysis is recommended.
Fortunately, the great majority of RSOs are characterized by orbits
that are well within the limits of validity of this theory.

\begin{table}[hbt!]
\fontsize{10}{10}\selectfont \caption{Initial osculating orbital elements of different test cases}
\label{NORAD_data} \centering %
\begin{tabular}{lccccc}
\hline 
RSO  & initial epoch (JD)  & a (km)  & e  & i (deg)  & $\omega$ (deg)\tabularnewline
\hline 
NORAD 47961  & 2,459,878.68  & 6,908.52  & 0.0025  & 97.50  & 65.67\tabularnewline
NORAD 41460  & 2,459,880.68  & 6,875.35  & 0.0125  & 98.27  & 129.61\tabularnewline
NORAD 43518  & 2,459,886.86  & 6,823.00  & 0.0006  & 34.93  & 289.07\tabularnewline
\hline 
\end{tabular}
\end{table}

\begin{figure}[t!]
\centering \subfloat[]{\includegraphics[width=0.49\textwidth]{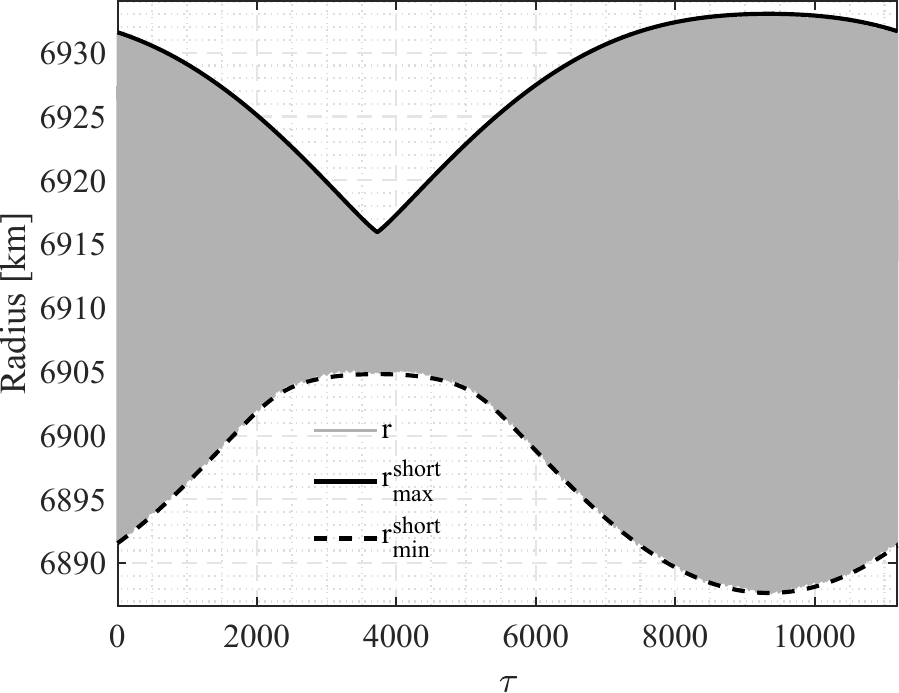}

}\subfloat[]{\includegraphics[width=0.49\textwidth]{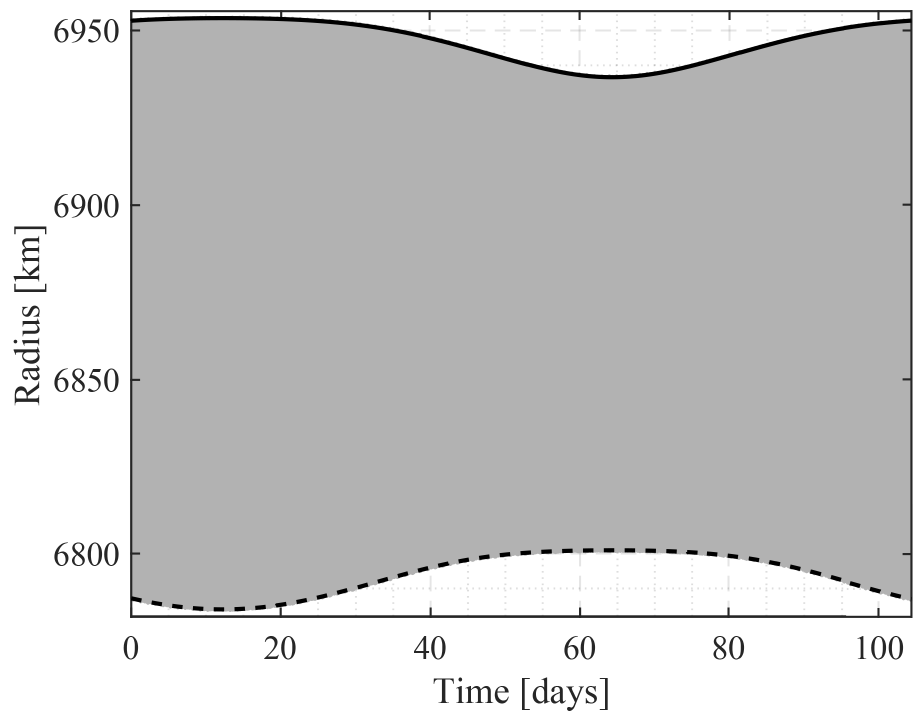}

}\\
 \subfloat[]{\includegraphics[width=0.49\textwidth]{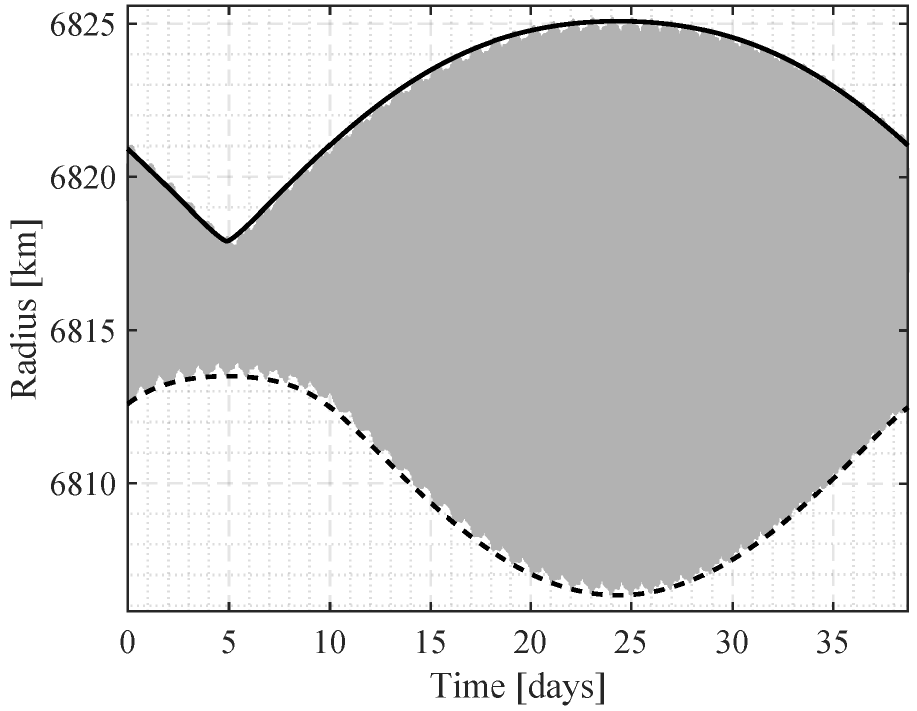}

}\caption{Short-term space occupancy bounding for NORAD 47961 (a), NORAD 41460
(b) and NORAD 43518 (c).}
\label{fig:envII} 
\end{figure}

Figure~\ref{fig:envII} compares the numerical propagated orbital
radius with the radial bounds obtained with the short-term SO model
and considering the three RSO test cases reported in Table \ref{NORAD_data}.
Note that the analytical computation only employs the initial osculating
orbital elements as input and goes through the following steps: 
\begin{enumerate}
\item The corresponding initial mean elements are computed with a Kozai-Lyddane
transformation (\cite{Kozai},\cite{Lyddane}) accounting for first
order J2 short-periodic components and reported in Appendix B. 
\item The value of $\hat{e}\left(\tau\right)$ , $\hat{\omega}\left(\tau\right)$,
$k$ are computed from Eqs. (\ref{eq:e_f}-\ref{eq:e_m2}). 
\item The maximum and minimum radius $\hat{r}$ corresponding to $\beta_{1}=\beta\left(\tau\right)$
is computed, after solving the quartic equation of Eq.(\ref{eq:SOquarticequation}),
from Eq. (\ref{eq:r_x}). 
\end{enumerate}
Figure~\ref{fig:envII} suggests that the qualitative behavior of
the long-period radial bounds fluctuations of different RSOs in LEO
can be well reproduced with the proposed analytical model.

In order to fully assess the validity of the model an extensive test
campaign has been conducted on a much wider number of test cases.
The considered dataset included 16,972 orbits, filtered to exclude
high eccentricity and high apogee orbits, and extracted from publicly
available Two-line Elements (TLEs) data and using SGP4 theory \cite{vallado2006revisiting}.

For each orbit the minimum and maximum radii provided by the numerical
propagation ($r_{max}^{num},r_{min}^{num}$) across a 5-day window
were compared to the corresponding values ($r_{max}^{short},r_{min}^{short}$)
of the analytical short-term SO model to compute the model error:

\[
\varepsilon=\max\left\{ \left|r_{max}^{short}-r_{max}^{num}\right|,\left|r_{min}^{short}-r_{min}^{num}\right|\right\}.
\]

\begin{figure}[t!]
\centering \includegraphics[width=0.49\textwidth]{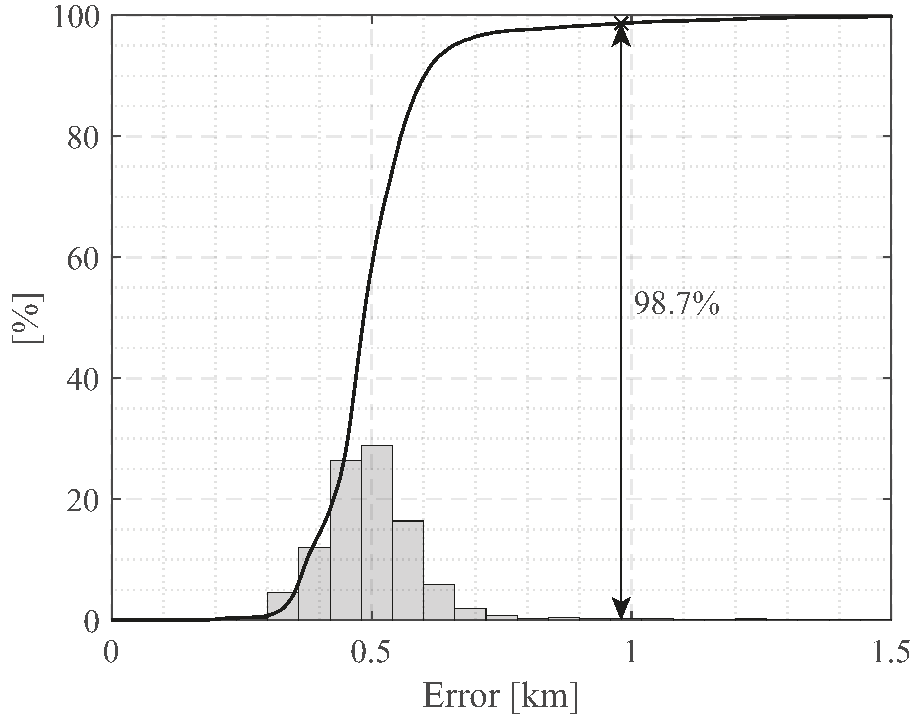}\caption{Short-term SO error histogram and cumulative density function.}
\label{fig:hist_STSO} 
\end{figure}

Figure~\ref{fig:hist_STSO} displays the error distribution and its
cumulative density function, revealing a mean error margin of approximately
0.5 kilometers. Given the inherent uncertainties in TLE-derived data,
this level of accuracy is deemed satisfactory. Remarkably, over 98.7\%
of the analyzed cases exhibited an error of less than one kilometer,
highlighting the robustness and applicability of the model.

\section{Space Occupancy Conjunction Filter}

\label{sec:filter}

The relatively good accuracy of the proposed SO model motivates its
application to first-stage conjunction filtering. More specifically,
it is attractive to consider an SO-based filter as a replacement of
the classical AP-filter, which typically serves as the initial filtering
stage in several conjunction analysis tools (see for instance \cite{smartsieve2})
and is based on the simple idea of eliminating pairs of objects with
non-overlapping ranges of radii.

The idea is to employ the short-term SO model as a more accurate,
yet computationally efficient tool in order to predict a radial overlap
for a given conjunction screening horizon. In addition to the classical
AP-filter \cite{CF}, three different implementations of a conjunction
filter based on the SO model are considered: 
\begin{enumerate}
\item ``AP-filter''. The classical filter where the radial bounds are
defined by apogee and perigee radius. 
\item ``SO-filter''. A filter implementation where the radial bounds are
evaluated through Eq.\eqref{eq:r_max_short}, which involves the solution
of the quartic equation, Eq. \eqref{eq:SOquarticequation}. 
\item ``SO-filter exact''. A refined but computationally expensive procedure
where the radial bounds are evaluated by solving Eq.~\eqref{partial_M}. 
\item ``SO-filter raw''. A very simple but crude implementation where
the maximum and minimum radii are obtained by bounding the two brackets
of Eq.~\eqref{eqn-hatr} by their respective extrema (as described
in \cite{SOfilter}). 
\end{enumerate}
To test and compare the filters, the same dataset and high-fidelity
model of Section~\ref{sec:validation} was employed. This set of
orbits provides a total of 144 million pairs to analyze. These elements
have been propagated to a common initial epoch $t_{0}$ (11/02/2022
09:18:20). The corresponding initial mean orbital elements have been
derived using Kozai-Lyddane equations\footnote{Note that implementing the filter based on the average orbital elements
extracted directly from the TLEs provided considerably worse performance.}.

A first comparison was conducted in terms of computational efficiency
showing that the SO-filter resulted in an increase in computation
time of only 1.8\% compared to the AP-filter while the SO-filter raw
was only slightly more efficient (with a 1.6\% increase). On the other
hand, the exact SO-filter showed a more substantial increase of 72.8\%.

Next, the different filters were compared in terms of accuracy and
effectiveness without applying any correction buffer in Section \ref{sec:without}.
Correction buffers (see \cite{BUF}) were then applied in order to
fully remove false negative errors and Section \ref{sec:with} provides
a comparative analysis of the performance of the modified filters.
Finally, Section \ref{sec:drag} describes the analysis of the effect
of atmospheric drag.

\subsection{Comparison of filters without buffer}

\label{sec:without}

A simple and preliminary assessment of the performance of the filters
has been conducted considering the same 16,972 RSOs dataset as in
Section \ref{sec:validation} over a 5-day time horizon and without
applying any correction buffer. The high-fidelity propagator described
in that section is employed to obtain the numerically propagated (``true'')
results in terms of radial bounds to be compared to the ones employed
by the different filters. Note that, for simplicity, the effect of
atmospheric drag has been neglected in this analysis and will be dealt
with in a dedicated section (Section~\ref{sec:drag}).

Following a widely adopted terminology, a \textit{positive} outcome
refers to a pair of orbits having overlapping ranges of radii, which
is a necessary but not sufficient condition for a collision to occur.
Conversely, a \textit{negative} outcome corresponds to non-overlapping
ranges of radii, which (in a deterministic analysis) rules out the
possibility of a collision within the established time window.

Given two orbits $i$ and $j$, each filter excludes the pair $(i,j)$
when at least one of the following two conditions is met: 
\begin{equation}
\begin{split}r_{max,i}<r_{min,j},\\
r_{max,j}<r_{min,i},
\end{split}
\label{eqn:filter-eqn-1}
\end{equation}
where $r_{max}$ and $r_{min}$ are the radial bounds computed by
a given filter and solely based on the state vectors of the two objects
at the reference time $t_{0}$. Specifically, for the AP-filter the
mean apogee and perigee at $t_{0}$ was employed since using osculating
values was seen to deteriorate the filter performance. This can be
appreciated in Fig.~\ref{radii} the osculating apogee and perigee
are highly time-dependent.

\begin{figure}[hbt!]
\centering\includegraphics[width=0.8\textwidth]{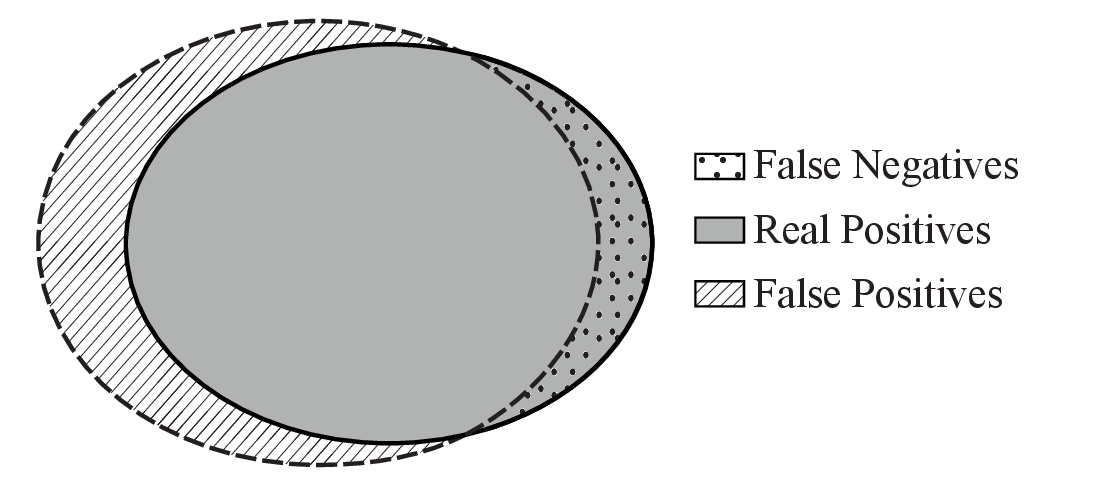}
\caption{Schematic of real (solid line) and filter-predicted (dash-line) positive
outcomes }
\label{FP_FN_schematic} 
\end{figure}

For a given filter, a \textit{false positive} error occurs whenever
the filter predicts a radial range overlap that does not take place
in that time window while a \textit{false negative} error occurs whenever
the filter fails to predict a radial range overlap and filters out
an object pair that could collide. A visual comparison of real vs.
filter-predicted positive outcomes is represented in Fig. \ref{FP_FN_schematic}.

From the total number of pairs ($N$), the number of pairs eliminated
by the filter ($N_{out}$), the number of real positive outcomes obtained
numerically ($N_{RP}$) and the number of false positives ($N_{FP})$
and false negatives ($N_{FN}$) scored by the filter, one can define
the \textit{false positives to detected real positives} ratio as:

\begin{equation}
\rho_{FP}=\frac{N_{FP}}{N_{RP}-N_{FN}},\label{eq:rho_FP}
\end{equation}
the \textit{false negatives to detected real positives} ratio as:

\begin{equation}
\rho_{FN}=\frac{N_{FN}}{N_{RP}-N_{FN}},\label{eq:rho_FN}
\end{equation}
and the \textit{filter effectiveness} as:

\begin{equation}
\eta=\frac{N_{out}}{N}.\label{eq:eta}
\end{equation}

Table~\ref{results_wob} summarizes the results of the four filters
using the three metrics previously defined.

\begin{table}[hbt!]
\fontsize{10}{10}\selectfont \caption{Summary of Filter Performance Comparison, without buffer.}
\label{results_wob} \centering %
\begin{tabular}{lcccccc}
\hline 
Filter  & Real Positives  & False Positives  & False Negatives  & $\rho_{FP}$  & $\rho_{FN}$  & $\eta$\tabularnewline
\hline 
AP-filter  & 31,950,589  & 224,312  & 537,201  & 0.714 \%  & 1.710 \%  & 77.659 \%\tabularnewline
SO-filter  & 31,950,589  & \textcolor{black}{2,310}  & \textcolor{black}{65,020}  & 0.007 \%  & 0.204 \%  & 77.813 \%\tabularnewline
SO-filter exact  & 31,950,589  & 1,889  & 67,054  & 0.006 \%  & 0.210 \%  & 77.813 \%\tabularnewline
SO-filter raw  & 31,950,589  & 445,342  & 6,752  & 1.394 \%  & 0.021 \%  & 77.505 \%\tabularnewline
\hline 
\end{tabular}
\end{table}

First of all, it should be noted that the less computationally expensive
SO-filter almost matches the performance of the SO-filter exact in
spite of the simplifying assumptions. In relation to the false positives,
the conservative character of the SO-filter raw makes it perform even
poorer than the AP-filter whereas the SO-filter manages to reduce
them by two orders of magnitude. On the other hand, compared to the
AP-filter, both the SO-filter and SO-filter raw considerably reduce
the number of false negatives.

\subsection{Comparison of filters with buffer}

\label{sec:with}

In conjunction analysis, false negative errors must be avoided since
they can leave potentially dangerous conjunctions undetected. To eliminate
false negatives, buffer distances are employed to ensure that a given
filter correctly bounds the radius evolution of each object in the
catalogue even at the expense of increasing false positives.

After adding buffer distances $b_{i}$ and $b_{j}$, a given filter
excludes from further conjunction assessment all pairs of satellites
obeying any of the two conditions: 
\begin{equation}
\begin{split}r_{max,i}+b_{i} & <r_{min,j}-b_{j},\\
r_{max,j}+b_{j} & <r_{min,i}-b_{i}.
\end{split}
\end{equation}

Ideally, for a given filter, the applied buffer size should be as
small as possible and with minimum sensitivity to the specific initial
conditions of the object considered.

In order to establish the appropriate size of a buffer for a given
filter, extensive testing needs to be conducted using the full catalogue
of objects and a high fidelity model as the one described in Section~\ref{sec:validation}.
As seen in Section~\ref{sec:validation}, the errors in computing
the radius bounds of an individual RSO vary considerably (depending
on the initial epoch) throughout one full line of apsides precession
period. This means that a conservative value needs to be computed
taking the maximum buffer size over that period\footnote{For those orbits for which this time interval exceeded one year the
evolution has been calculated over 365 days}.

\begin{figure}[hbt!]
\centering \includegraphics[width=0.49\textwidth]{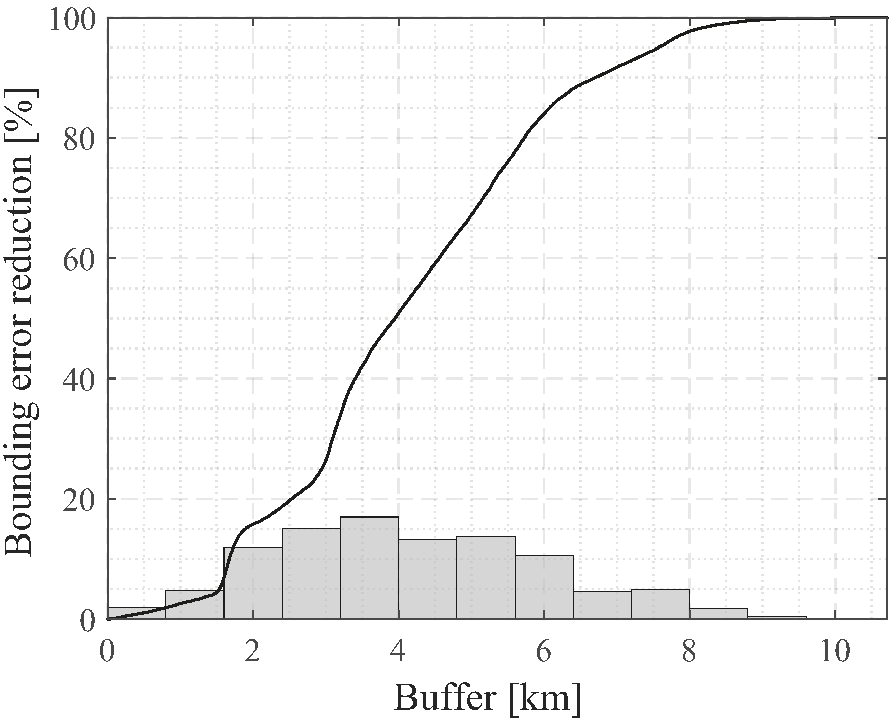}\includegraphics[width=0.49\textwidth]{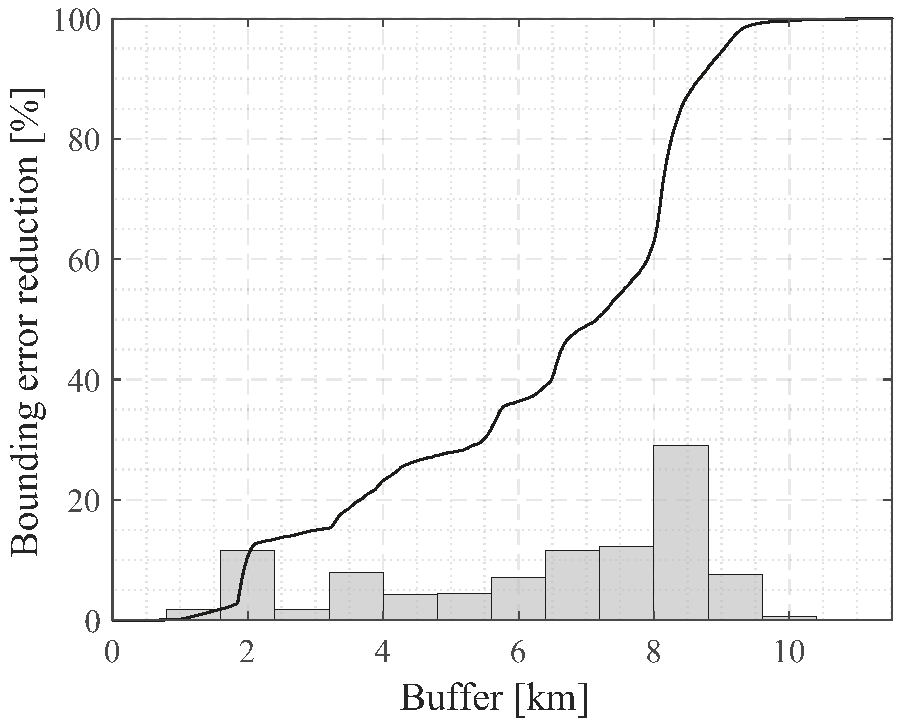}

\includegraphics[width=0.49\textwidth]{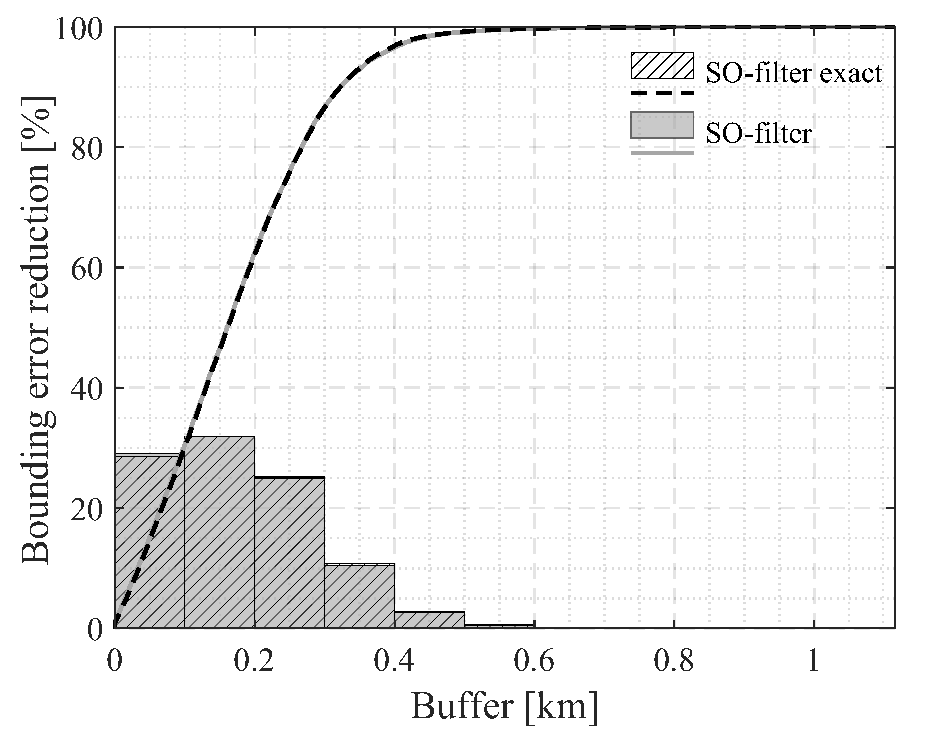}\includegraphics[width=0.49\textwidth]{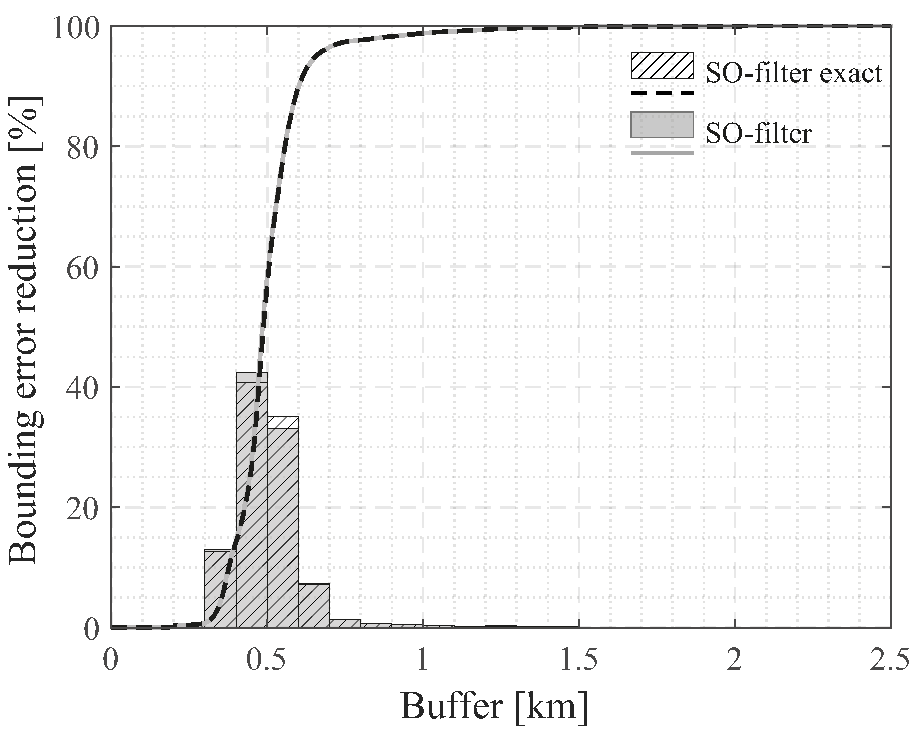}
\caption{Statistics of the bounding error reduction as a function of the applied
buffer for the AP-filter (top) and the SO-filter (bottom) and for
a 5-day (left) and full precession period time span (right).}
\label{hist_cdf_ini} 
\end{figure}

Figure~\ref{hist_cdf_ini} presents the statistics of the bounding
error reduction as a function of the applied buffer for each of the
four filters considered and using the same 16,972 RSOs population.
The results displayed in the left-hand side of the figure take into
account a 5-day time span while the ones in the right-hand side are
based on a full line of apsides precession period (a few months).
The reference initial epoch is $JD=2,459,885.89$ for both cases.
It can be seen that the behavior is similar; however, extending the
propagation over a precession period moves the histogram to the right
thereby increasing the buffers. The plots show a typical 9 km required
buffer for the AP-filter (3-6 km for the 5-day time span) compared
to a 0.5 km for the SO-filters (0.1-0.2 km for the 5-day time span).
A complete elimination of bounding error for the AP-filter requires
a buffer of 11.527 km (10.722 km for the 5-day time span) while the
SO-filter requires a buffer of 2.508 km (1.119 km for the 5-day time
span).

The preceding results suggest a possible improvement in terms of buffer
minimization for the SO-filter. The fact that the maximum required
buffer is five times greater than the statistical mode of its distribution
leads to the consideration of whether the few cases requiring a larger
buffer share some common characteristics and may be treated differently.

Indeed, it has been observed that these few cases are characterized
by more eccentric and higher altitude orbits. Therefore, in order
to avoid penalizing the entire population with such a high buffer,
it has been decided to divide the population into six distinct groups
based on the minimum altitude and eccentricity and apply a bespoke
buffer to each category for each filter. Regarding eccentricity, the
population has been divided into two groups. Within the low eccentricity
orbits, four categories have been distinguished based on minimum initial
mean altitude: the first one with a minimum altitude below 400 km,
the second group with a minimum altitude between 400 and 700 km, the
third group between 700 and 1,000 km, and finally, higher altitude
orbits, including all those above 1,000 km. On the other hand, orbits
with higher eccentricities have been divided into two groups: low-altitude
orbits with a minimum altitude below 1,000 km, and high-altitude orbits
comprising all others. Table~\ref{tab:buf_cat} presents the required
buffer for each filter and each category.

\begin{table}[hbt!]
\fontsize{10}{10}\selectfont \caption{Required buffer for each filter according to categories.}
\label{tab:buf_cat} \centering %
\begin{tabular}{llcccc}
\hline 
Eccentricity  & Altitude {[}km{]}  & SO-filter {[}km{]}  & AP-filter {[}km{]}  & SO-filter exact {[}km{]}  & SOS-filter {[}km{]}\tabularnewline
\hline 
$e<0.01$  & $r_{min}-R_{\oplus}<400$  & 0.9782  & 11.5271  & 0.9780  & 1.2586\tabularnewline
 & $r_{min}-R_{\oplus}<700$  & 1.2823  & 11.2849  & 1.2827  & 3.3474\tabularnewline
 & $r_{min}-R_{\oplus}<1,000$  & 0.7066  & 10.2531  & 0.7066  & 3.0379\tabularnewline
 & $r_{min}-R_{\oplus}>1,000$  & 2.0260  & 8.5749  & 2.0263  & 2.8355\tabularnewline
\hline 
$0.1<e<0.01$  & $r_{min}-R_{\oplus}<1,000$  & 0.9009  & 10.7209  & 0.9018  & 3.0047\tabularnewline
 & $r_{min}-R_{\oplus}>1,000$  & 2.5072  & 8.4504  & 2.5076  & 2.7253\tabularnewline
\hline 
\end{tabular}
\end{table}

Table~\ref{tab:final_results} summarizes the results of the four
filters with their buffers presented in Table~\ref{tab:buf_cat}.
It is important to highlight that this latest improvement represents
a decrease of more than 50\% in false positives for the SO-filter,
removing almost 800,000 more pairs at a very small additional cost
when compared to the case of a common applied buffer.

\begin{table}[hbt!]
\fontsize{10}{10}\selectfont \caption{Summary of Filter Performance Comparison with the required buffer.}
\label{tab:final_results} \centering %
\begin{tabular}{lccccc}
\hline 
Filter  & Real Positives  & False Positives  & False Negatives  & $\rho_{FP}$  & $\eta$\tabularnewline
\hline 
AP-filter  & 31,950,589  & 5,507,984  & 0  & 17.239 \%  & 73.990 \%\tabularnewline
SO-filter  & 31,950,589  & \textcolor{black}{530,720}  & 0  & 1.661 \%  & 77.446 \%\tabularnewline
SO-filter exact  & 31,950,589  & 524,183  & 0  & 1.641 \%  & 77.451 \%\tabularnewline
SO-filter raw  & 31,950,589  & 2,145,317  & 0  & 6.714 \%  & 76.325 \%\tabularnewline
\hline 
\end{tabular}
\end{table}

To close the comparison, the behavior of the $\rho_{FP}$ and $\rho_{FN}$
ratios with respect to the buffer size applied to the SO-filter and
the AP-filter has been analyzed in a similar way as done in \cite{BUF}.

\begin{figure}[hbt!]
\centering \includegraphics[width=0.7\textwidth]{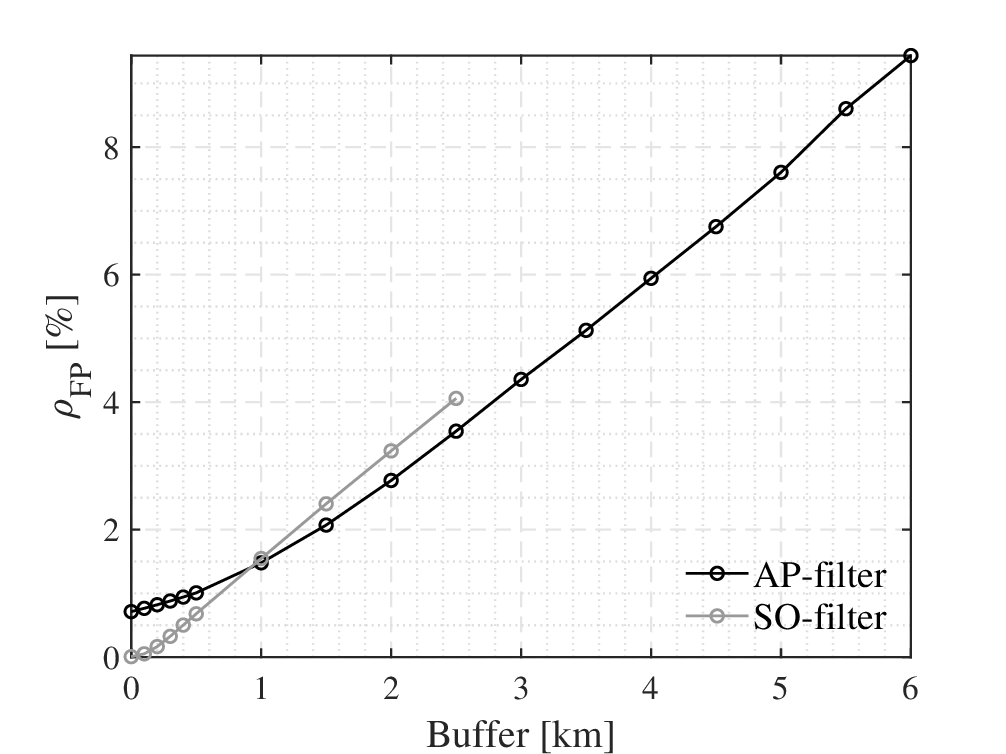}
\caption{False positive to detected real positive ratio as a function of the
applied buffer}
\label{fig:TypeI} 
\end{figure}

Figure~\ref{fig:TypeI} illustrates the relationship between the
$\rho_{FP}$ and buffer size. As expected, the false positive ratio
increases with larger buffer sizes. Interestingly, the SO-filter outperforms
the AP-filter for buffer sizes below 1 km, but beyond this threshold,
the AP-filter demonstrates superior performance.

\begin{figure}[hbt!]
\centering \includegraphics[width=0.7\textwidth]{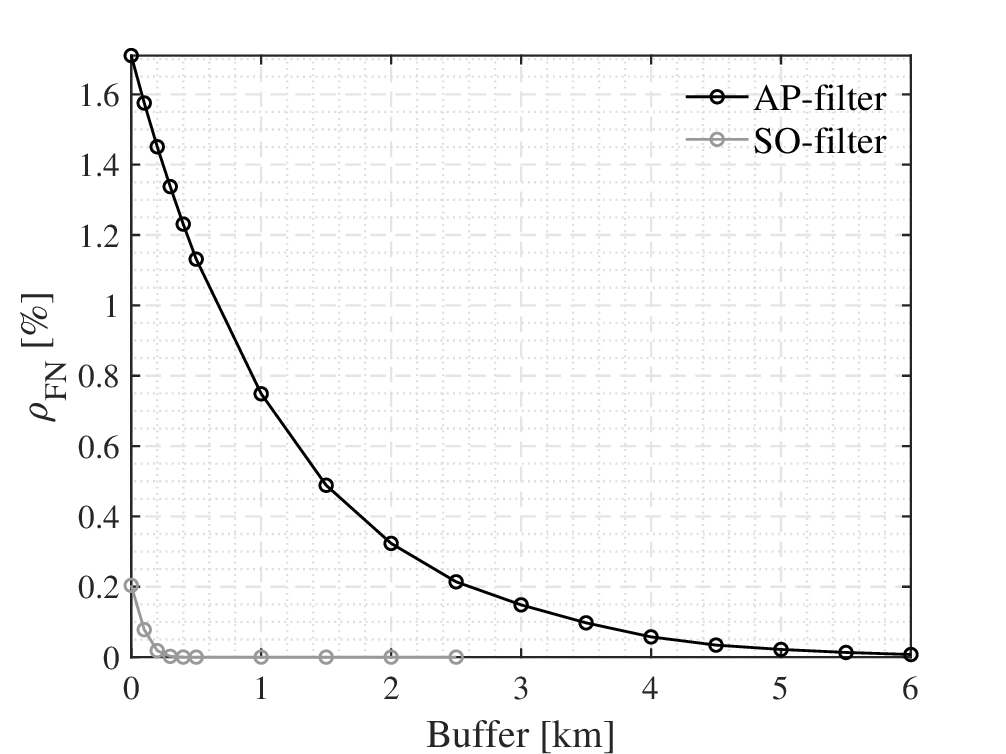}
\caption{False positive to detected real positive ratio as a function of the
applied buffer}
\label{fig:TypeII} 
\end{figure}

Figure~\ref{fig:TypeII} plots the $\rho_{FN}$ against the buffer.
This graph demonstrates the superior performance of the SO filter.
Notably, a buffer of 0.5 km is sufficient to eliminate all false negatives
when using the SO-filter, whereas the AP-filter necessitates a buffer
exceeding 6 km to achieve the same outcome. 

\begin{figure}[hbt!]
\centering \includegraphics[width=0.7\textwidth]{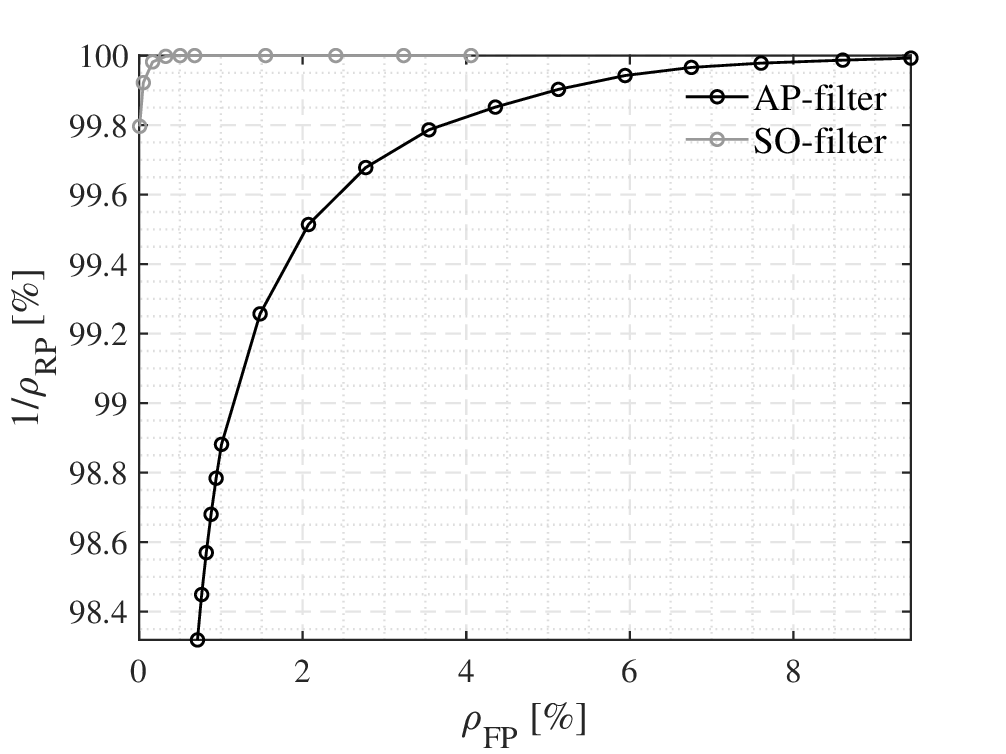}
\caption{Operating characteristic curve.}
\label{fig:OpCh} 
\end{figure}

Finally, Fig.~\ref{fig:OpCh} plots the fraction of detected real
positives:

\begin{equation}
\frac{1}{\rho_{RP}}=\frac{N_{RP}-N_{FN}}{N_{RP}}=\frac{1}{1+\rho_{FN}},\label{eq:rho_RP}
\end{equation}
against the previously defined false positive to detected real positive
rate $\rho_{FP}$ (Eq \eqref{eq:rho_FP}) showing what is known as
receiver operating characteristic curve \cite{ROC}. This type of
curve is used to compare two operating characteristics and depicts
relative trade-offs between benefits (true positives) and costs (false
positives). The filter performance, which is higher the closer this
curve is to the upper left corner, is clearly much higher for the
SO-filter case than the classical AP-filter.

\subsection{Atmospheric drag}

\label{sec:drag} The previous results do not take into account the
effect of atmospheric drag, which is relevant only for relatively
low altitude orbits (say, below 500 km). In order to extend the SO
analysis and filter design to include this effect what can be done
is to add a lower altitude buffer to account for the drag-induced
decay with a simplified model that allows a quick estimation to be
included in the filter estimates. A simple recipe to do that is to
exploit the starred ballistic coefficient ($B^{*}$) publicly available
from TLEs data. The following analysis investigates this possibility
to extend the SO model accordingly, and computes the updated filter
performance metrics.

The decrease in minimum altitude as a function of time for a near
circular orbit can be estimated from Eq.~\eqref{eq:ht} developed
by Billik \cite{drag}:

\begin{equation}
h=\frac{1}{\beta_{a}}\log\left[e^{\beta_{a}h_{0}}-B\sqrt{\mu R_{\oplus}}\beta_{a}\bar{\rho}t\right],\label{eq:ht}
\end{equation}
where $B$ is the ballistic coefficient, $h_{0}$ is the initial altitude,
$t$ is time, and $\beta_{a}$, $\bar{\rho}$ are coefficients characterizing
an exponential density model according to:

\begin{equation}
\rho=\bar{\rho}e^{-\beta_{a}h}.\label{eq:ctes}
\end{equation}

The classical definition for the ballistic coefficient is given in
terms of the drag coefficient of the satellite $C_{D}$, the frontal
area $A_{v}$ and the satellite mass, $m_{v}$ as:

\begin{equation}
B=\frac{C_{D}A_{v}}{m_{v}}.\label{bc}
\end{equation}

From Vallado~\cite{vallado2001fundamentals}, the classical ballistic
coefficient can be related to the starred ballistic coefficient readily
available from TLEs as:

\begin{equation}
B=12.741621B^{*}.\label{B_const}
\end{equation}

The computation of the two coefficients of the exponential density
model was carried out using the procedure explained in \cite{drag}.
This involves substituting two pairs of data, altitude, and density,
into Eq. \eqref{eq:ctes} and solving the system to obtain the values
of the coefficients. A layered model approach was employed, computing
$\beta_{a}$ and $\bar{\rho}$ for each 50 km altitude layer. Altitude
and density data were obtained from the exponential model in \cite{vallado2001fundamentals},
utilizing the data from Table 8-4. Table \ref{tab:fit_drag} displays
the coefficient values for each layer.

\begin{table}[hbt!]
\fontsize{10}{10}\selectfont \caption{Coefficients characterizing the exponential density model.}
\label{tab:fit_drag} \centering %
\begin{tabular}{lcc}
\hline 
Minimum altitude {[}km{]}  & $\beta_{a}$ {[}1/km{]}  & $\bar{\rho}$ {[}km/m$^{3}${]}\tabularnewline
\hline 
$r_{min}-R_{\oplus}<175$  & 0.0549  & $8.059\times10^{-06}$ \tabularnewline
$r_{min}-R_{\oplus}<225$  & 0.0404  & $6.426\times10^{-07}$ \tabularnewline
$r_{min}-R_{\oplus}<275$  & 0.0220  & $1.013\times10^{-08}$ \tabularnewline
$r_{min}-R_{\oplus}<325$  & 0.0186  & $4.078\times10^{-09}$ \tabularnewline
$r_{min}-R_{\oplus}<375$  & 0.0195  & $5.440\times10^{-09}$ \tabularnewline
$r_{min}-R_{\oplus}<425$  & 0.0163  & $1.629\times10^{-09}$ \tabularnewline
$r_{min}-R_{\oplus}<500$  & 0.0164  & $1.716\times10^{-09}$ \tabularnewline
\hline 
\end{tabular}
\end{table}

\begin{figure}[hbt!]
\centering \includegraphics[width=0.7\textwidth]{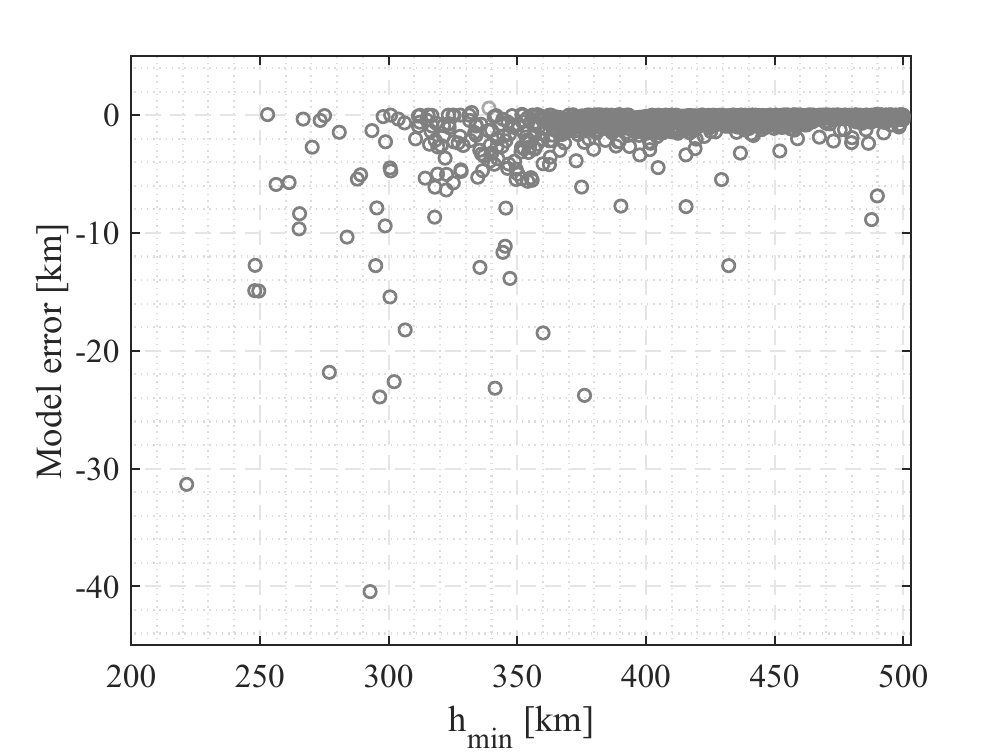}
\caption{Model error with respect to the minimum real altitude.}
\label{fig:model_error} 
\end{figure}

Before incorporating Eq.(\ref{eq:ht}) into the filter a numerical
test campaign has been conducted to verify that it is sufficiently
conservative. The analytically computed altitude drop has been compared
with the results of a numerical integration based on the high-fidelity
model of Section~\ref{sec:validation} now including the atmospheric
drag (using the exponential density model presented in \cite{vallado2001fundamentals},
for consistency). The same dataset of 16,972 RSOs orbits considered
in Section \ref{sec:validation} has been numerically propagated over
a 5-day timespan and considering the B{*} drag coefficient (Eq.~\eqref{B_const})
as provided in the TLEs catalog. The decrease in altitude predicted
by the analytical model is compared with the numerically propagated
one . Figure~\ref{fig:model_error} summarizes the difference between
the two (excluding reentered objects, i.e. with altitudes reaching
below 150 km). It can be seen that the model overestimates the descent
in the great majority of cases. In order to make the filter fully
conservative, a safety margin of 0.6 km has been added to ensure that
the bounds used cover the entire space occupancy of the orbits, avoiding
false negatives. In those cases where the model predicts reentry,
the minimum radius has been set to 0, since the study of reentry is
out of the scope of this analysis.

In order to analyze and compare the performance of the SO-filter with
the added atmospheric drag correction the results of Table \ref{tab:final_results}
have been recomputed. Note that a drag correction is only considered
for RSOs below 500 km altitude. Table \ref{tab:results_drag} summarizes
the new results.

\begin{table}[hbt!]
\fontsize{10}{10}\selectfont \caption{Summary of Filter Performance Comparison including atmospheric drag.}
\label{tab:results_drag} \centering %
\begin{tabular}{lcccccc}
\hline 
Filter  & Real Positives  & False Positives  & False Negatives  & $\rho_{FP}$  & $\rho_{FN}$  & $\eta$\tabularnewline
\hline 
AP-filter  & 31,950,589  & 5,520,894  & 0  & 17.279 \%  & 0 \%  & 73.981 \%\tabularnewline
SO-filter  & 31,950,589  & \textcolor{black}{536,507}  & 0  & 1.679 \%  & 0 \%  & 77.442 \%\tabularnewline
SO-filter exact  & 31,950,589  & 537,930  & 0  & 1.684 \%  & 0 \%  & 77.441 \%\tabularnewline
SO-filter raw  & 31,950,589  & 2,158,894  & 0  & 6.757 \%  & 0 \%  & 76.315 \%\tabularnewline
\hline 
\end{tabular}
\end{table}

It can be seen that the $\rho_{FP}$ increases slightly compared to
the previous analysis, by only 0.03-0.04\%. Additionally, it is evident
that the atmospheric drag affects the same way the four filters confirming
the conclusions of Section~\ref{sec:with}.

\section{Space Traffic Management Analysis}

\label{sec:evolution}

As shown in this section, the field of application of the tools developed
in this work is not limited to conjunction filtering alone and can
be extended to space traffic management (STM) analysis. From an STM
perspective, it is important to identify the total number of object
pairs occupying overlapping orbital shells in the past and current
space environment. In addition, for a given spacecraft, it is interesting
to assess how many \textsl{neighboring} RSOs (i.e., objects with overlapping
space occupancy) it has.This can provide a quick and useful information
on the degree by which a given object orbit is transecting through
the surrounding space environment.

By applying the SO-filter to historical space object data, one can
obtain insights into the changing dynamics of the orbital environment
and evaluate the performance of space traffic management efforts over
time. The dataset for this analysis was obtained from the Space Track
cloud storage site considering the total number of objects in the
catalog on January 1 of each year.

\begin{figure}[hbt!]
\centering \includegraphics[width=0.7\textwidth]{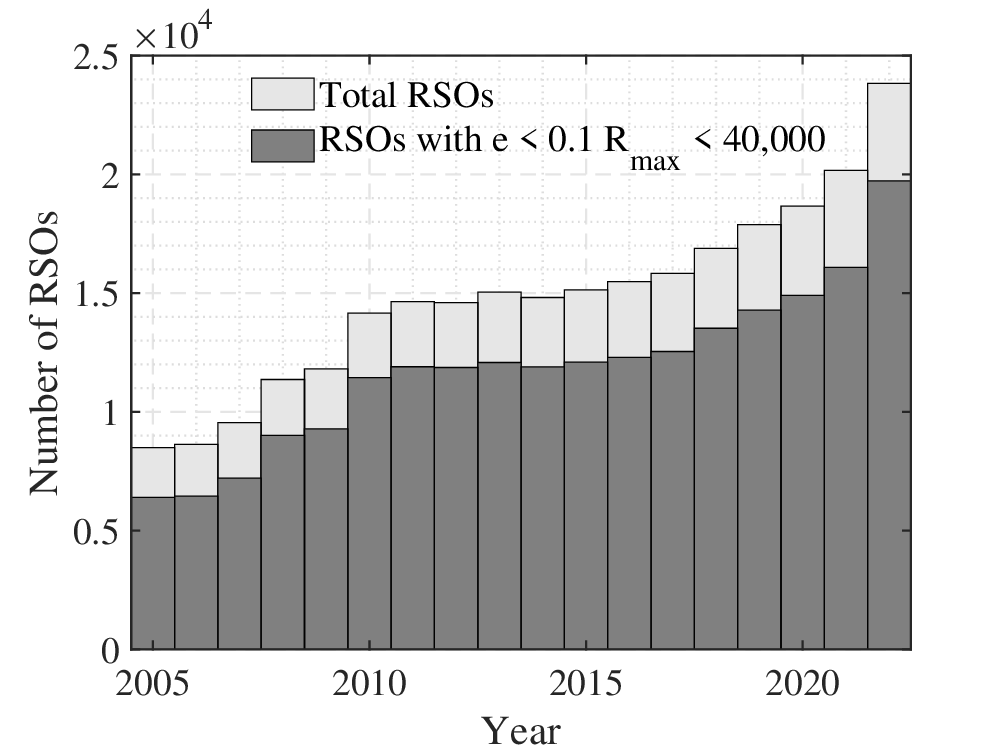}
\caption{Evolution of the space population from 2005 to the present.}
\label{fig:evolution} 
\end{figure}

\begin{figure}[hbt!]
\centering \includegraphics[width=0.7\textwidth]{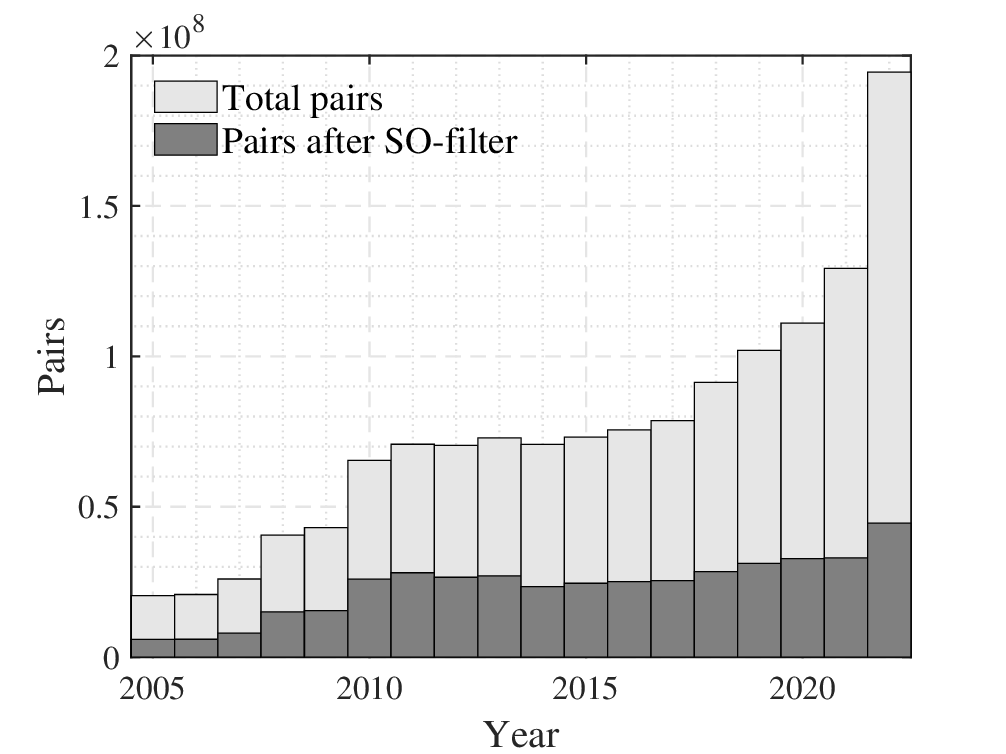}
\caption{Evolution of space distribution from 2005 to the present.}
\label{fig:SO_sh} 
\end{figure}

Figure~\ref{fig:evolution} illustrates the rapid growth in the total
number of objects in space over the last two decades, as well as the
increase in the population of objects to which the SO-filter can be
applied (orbits with apogee radii below approximately 40,000 km and
eccentricities below 0.1, as determined in Section \ref{sec:validation}).
Interestingly, while the total number of objects and potential collision
pairs has steadily increased, Fig.~\ref{fig:SO_sh} reveals that
the number of pairs sharing space occupancy remained relatively constant
between 2010 and 2021. In 2010, 40\% of object pairs shared space
occupancy, decreasing to 26\% in 2021 and 23\% in 2022. This suggests
improved space traffic management in recent years, although the sharp
increase in shared space occupancy pairs in the last year, primarily
due to the launch of new megaconstellations, warrants attention.

\begin{figure}[hbt!]
\centering \includegraphics[width=0.49\textwidth]{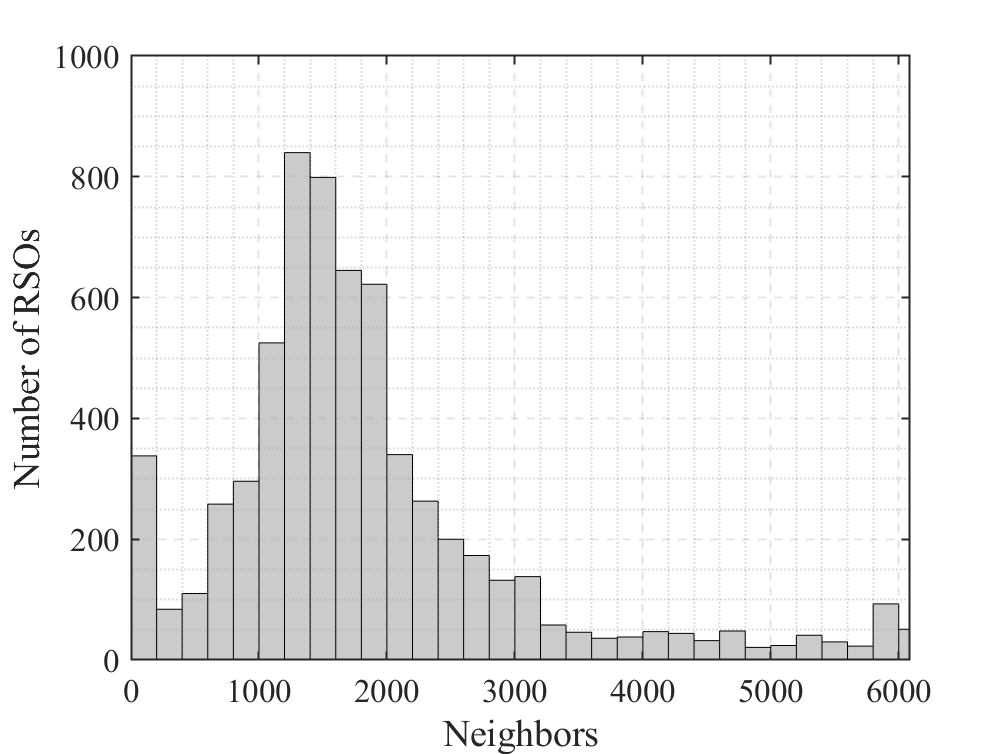}
\includegraphics[width=0.49\textwidth]{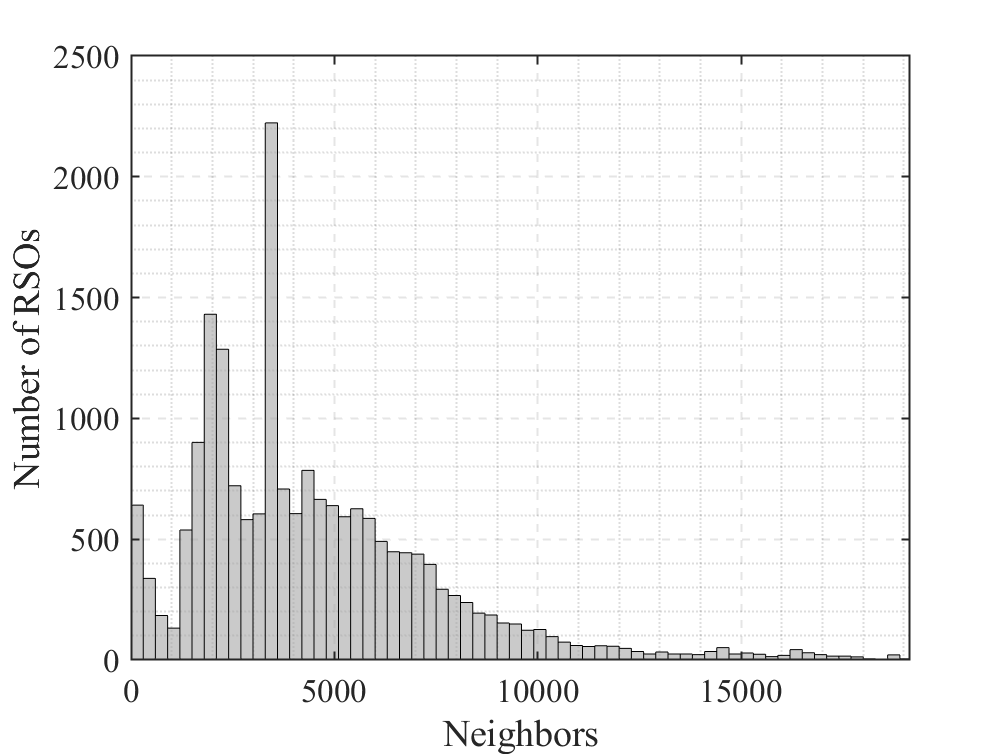} \caption{Neighbors histograms in 2005 (left) and 2022 (right).}
\label{fig:Nei_hist} 
\end{figure}

To further investigate this trend, the number of neighboring objects-{}-{}-those
sharing space occupancy with a given object-{}-{}-was examined.
Figure~\ref{fig:Nei_hist} presents histograms comparing the distribution
of neighborhood population sizes in 2005 and 2022. The 2.5-fold increase
in the number of cataloged space objects during this period naturally
led to an increase in the number of neighbors per object orbit. In
2005, most objects had around 1,500 neighbors, whereas in 2022, the
majority had between 2,000-4,000 neighbors, with the largest neighborhoods
tripling in size.

\begin{figure}[hbt!]
\centering \includegraphics[width=0.49\textwidth]{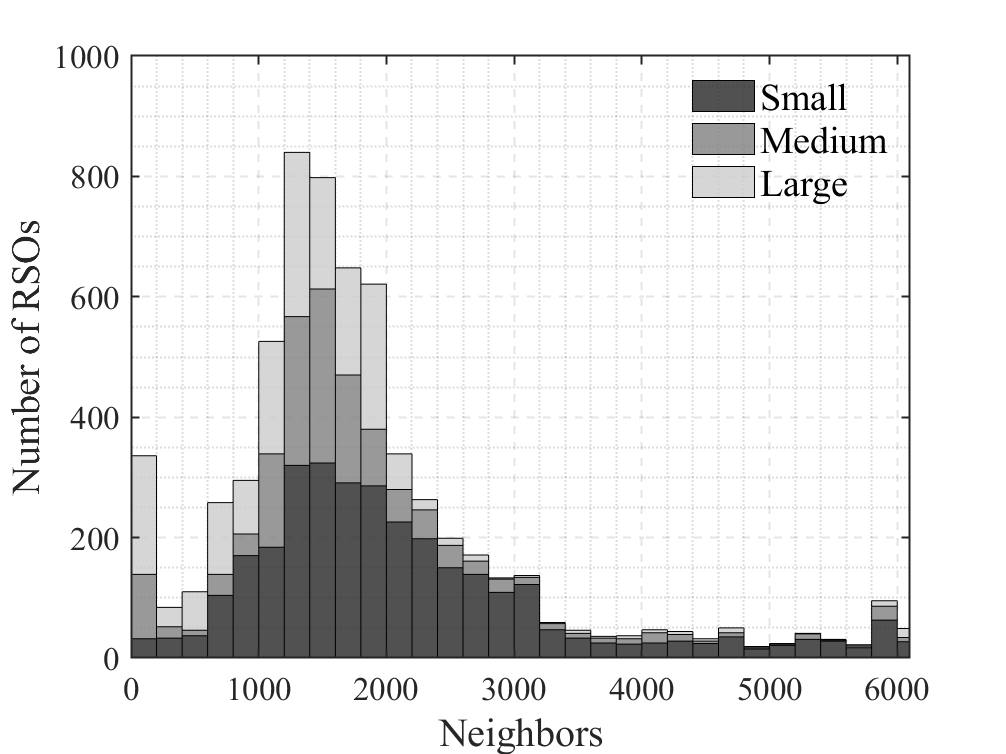}
\includegraphics[width=0.49\textwidth]{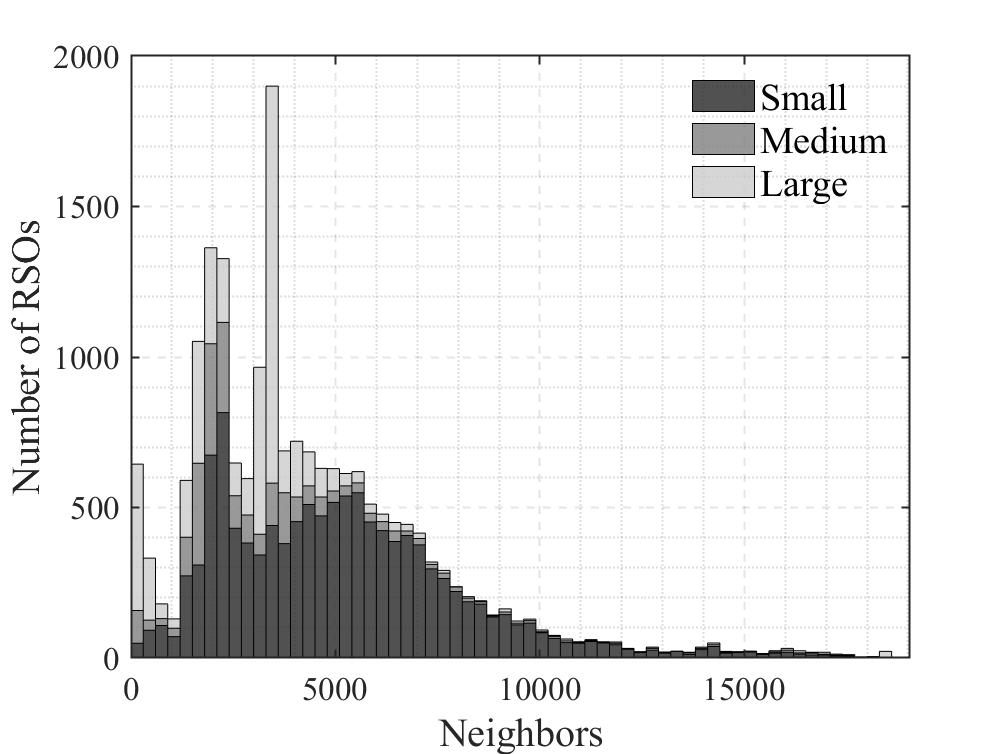}
\caption{Neighbors histograms in 2005 (left) and 2022 (right) depending on
main object size.}
\label{fig:Nei_hist_size} 
\end{figure}

Considering the impact of satellite size on collision risk, the objects
were categorized as small, medium, or large based on Space Track data.
Figure~\ref{fig:Nei_hist_size} displays the neighborhood size histograms
for 2005 and 2022, distinguishing the main object size in each neighborhood.
In both years, small objects generally had the largest neighborhoods,
while medium and large objects tended to have smaller neighborhoods.
However, the number of small objects grew considerably more than the
medium and large object populations by 2022.

\begin{figure}[hbt!]
\centering \includegraphics[width=0.49\textwidth]{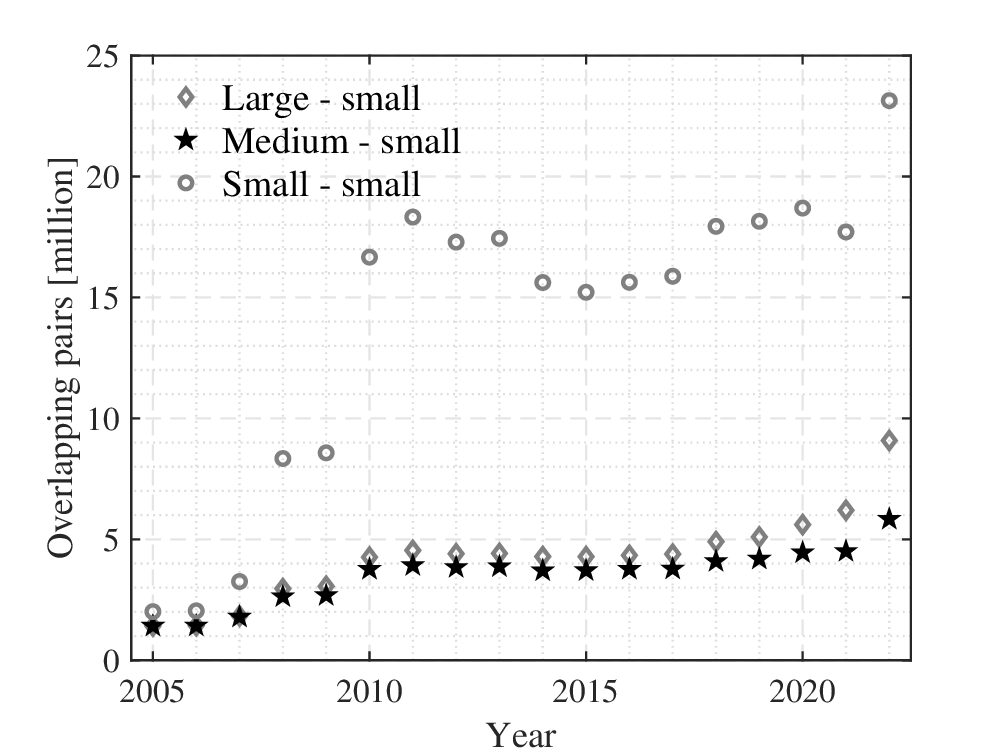}
\includegraphics[width=0.49\textwidth]{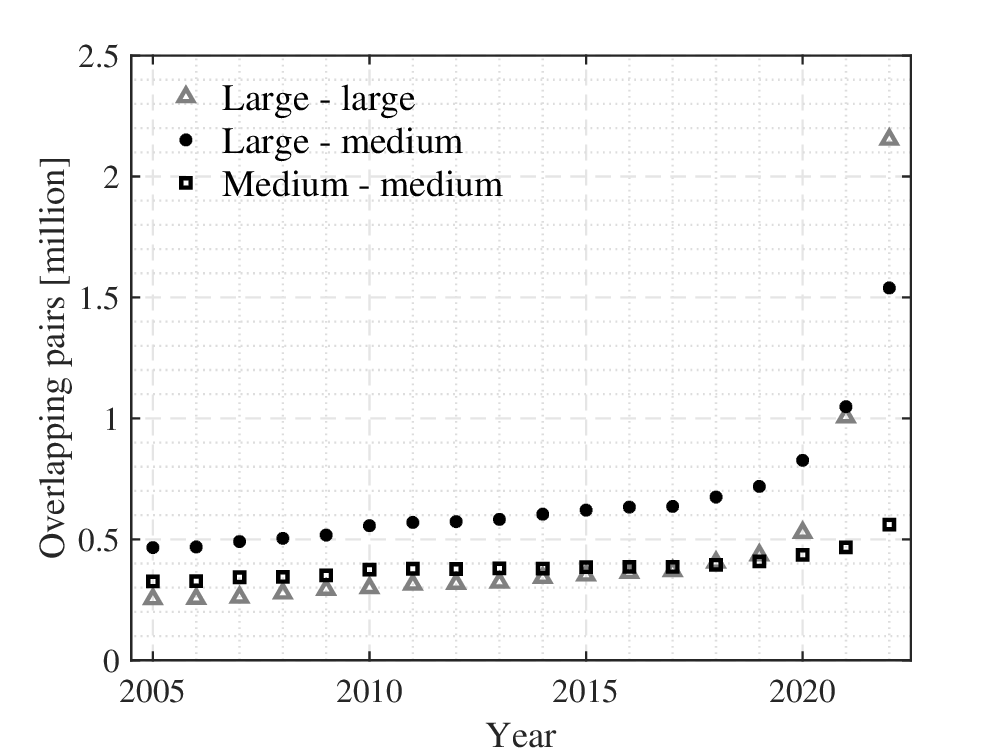}\caption{Evolution of object overlapping pairs distribution from 2005 to the
present by object size.}
\label{fig:size_pairs} 
\end{figure}

To further investigate collision risk, object pairs were classified
into six categories based on the sizes of both objects involved, as
collisions between larger objects pose a greater threat to the environment.
Figure~\ref{fig:size_pairs} shows the evolution of pairs sharing
space occupancy according to this classification. Pairs involving
small objects represent the majority and exhibit the greatest growth
over the years, while pairs consisting solely of medium or large objects
are less common and remained fairly constant until a slight increase
in the last two years.

\begin{figure}[hbt!]
\centering \includegraphics[width=1\textwidth]{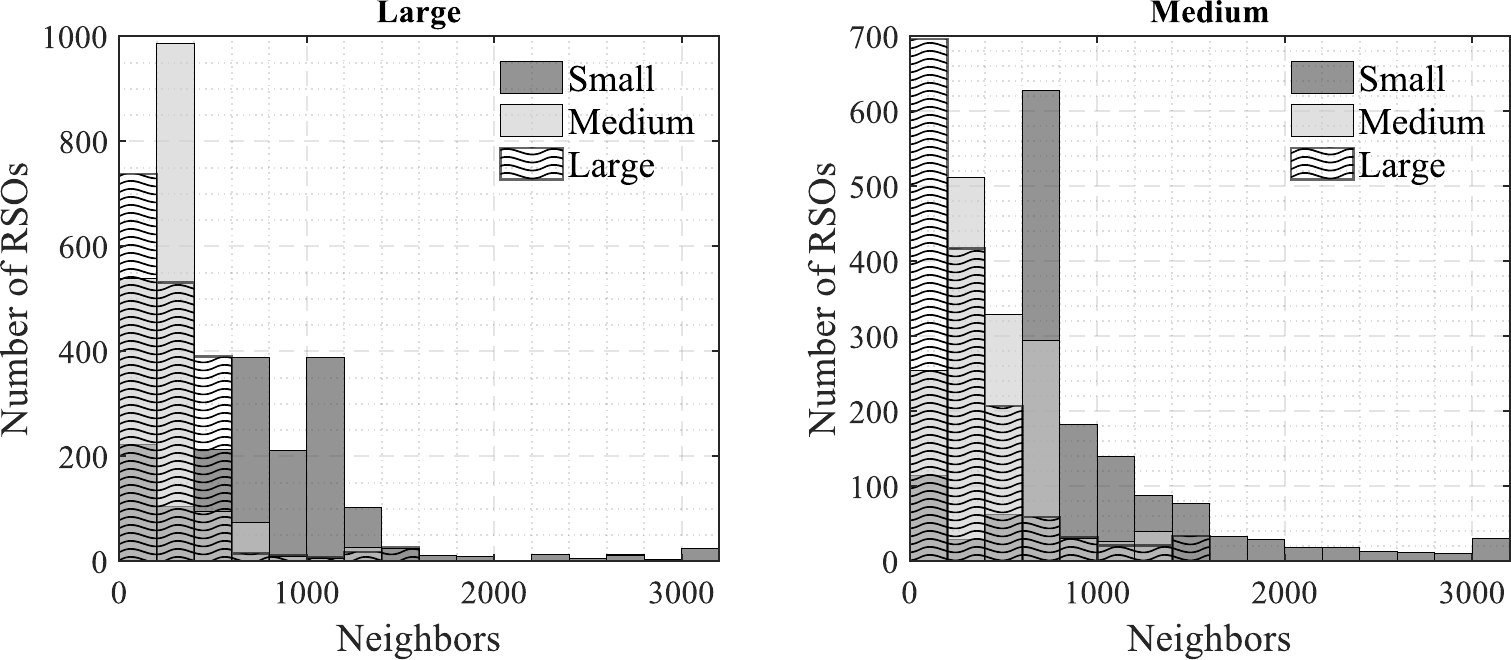}
\caption{Neighbors histograms of large (left) and medium (right) main objects
in 2005.}
\label{hist_05} 
\end{figure}

\begin{figure}[hbt!]
\centering \includegraphics[width=1\textwidth]{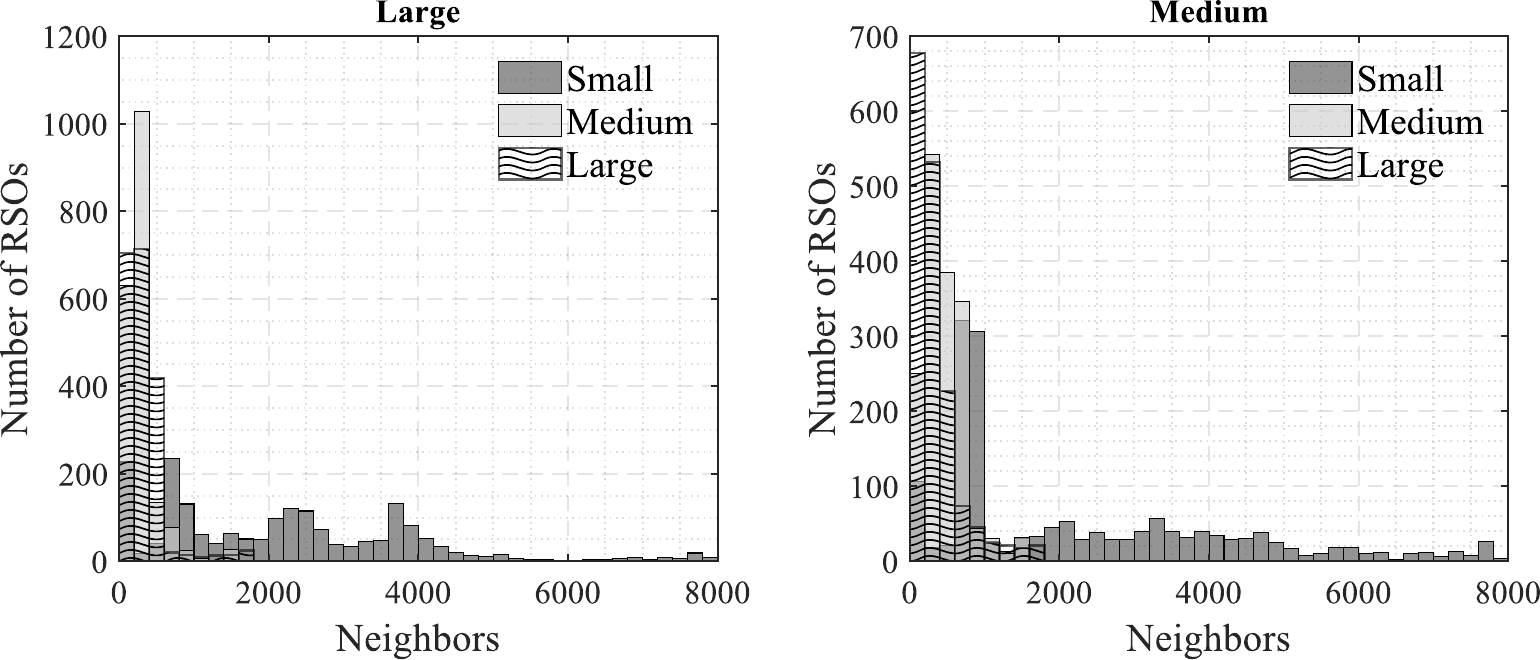}
\caption{Neighbors histograms of large (left) and medium (right) main objects
in 2010.}
\label{hist_10} 
\end{figure}

\begin{figure}[hbt!]
\centering \includegraphics[width=1\textwidth]{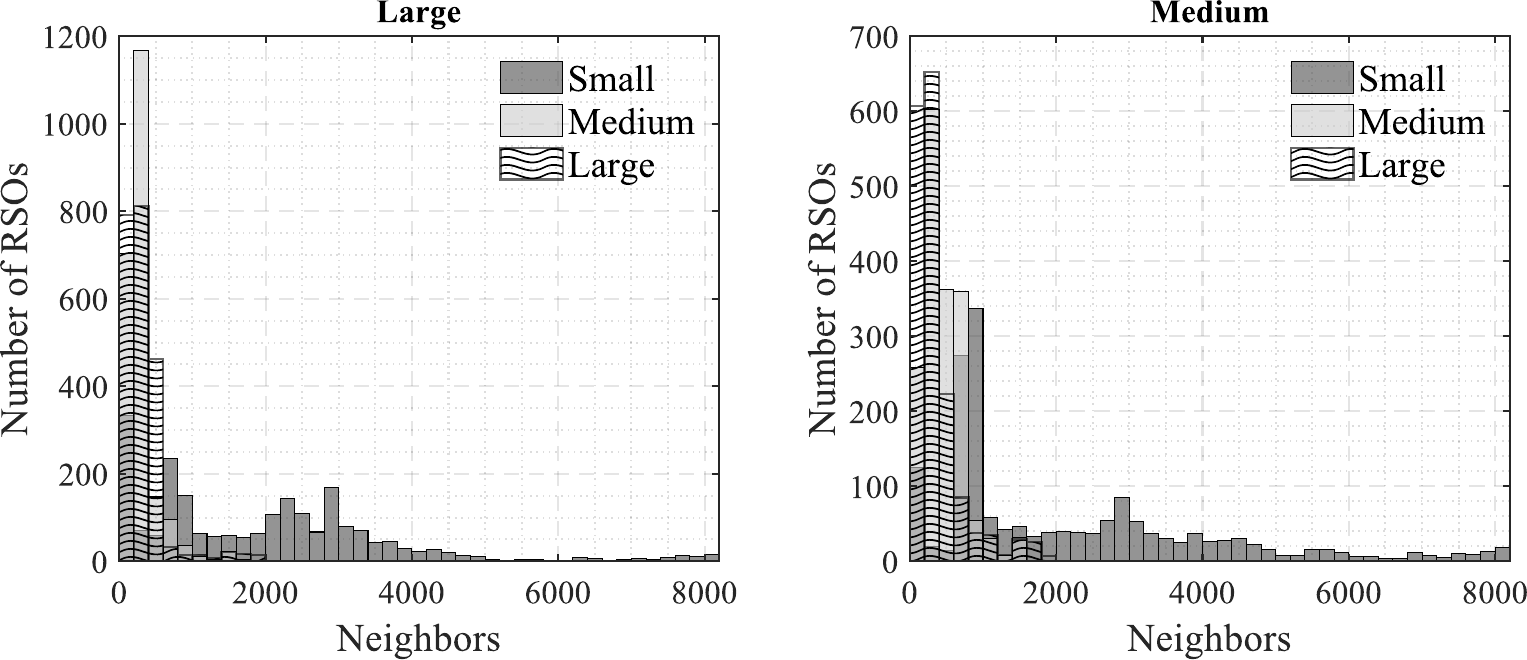}
\caption{Neighbors histograms of large (left) and medium (right) main objects
in 2015.}
\label{hist_15} 
\end{figure}

\begin{figure}[hbt!]
\centering \includegraphics[width=1\textwidth]{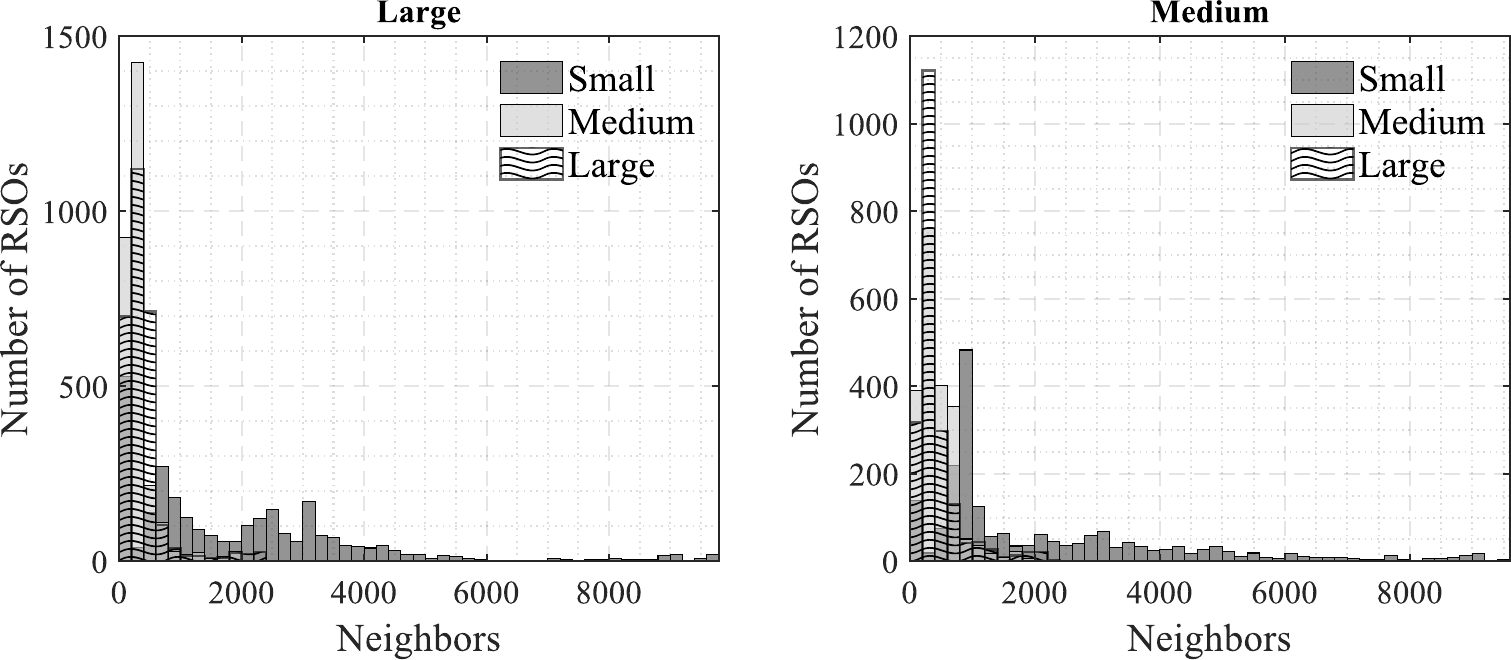}
\caption{Neighbors histograms of large (left) and medium (right) main objects
in 2020.}
\label{hist_20} 
\end{figure}

\begin{figure}[hbt!]
\centering \includegraphics[width=1\textwidth]{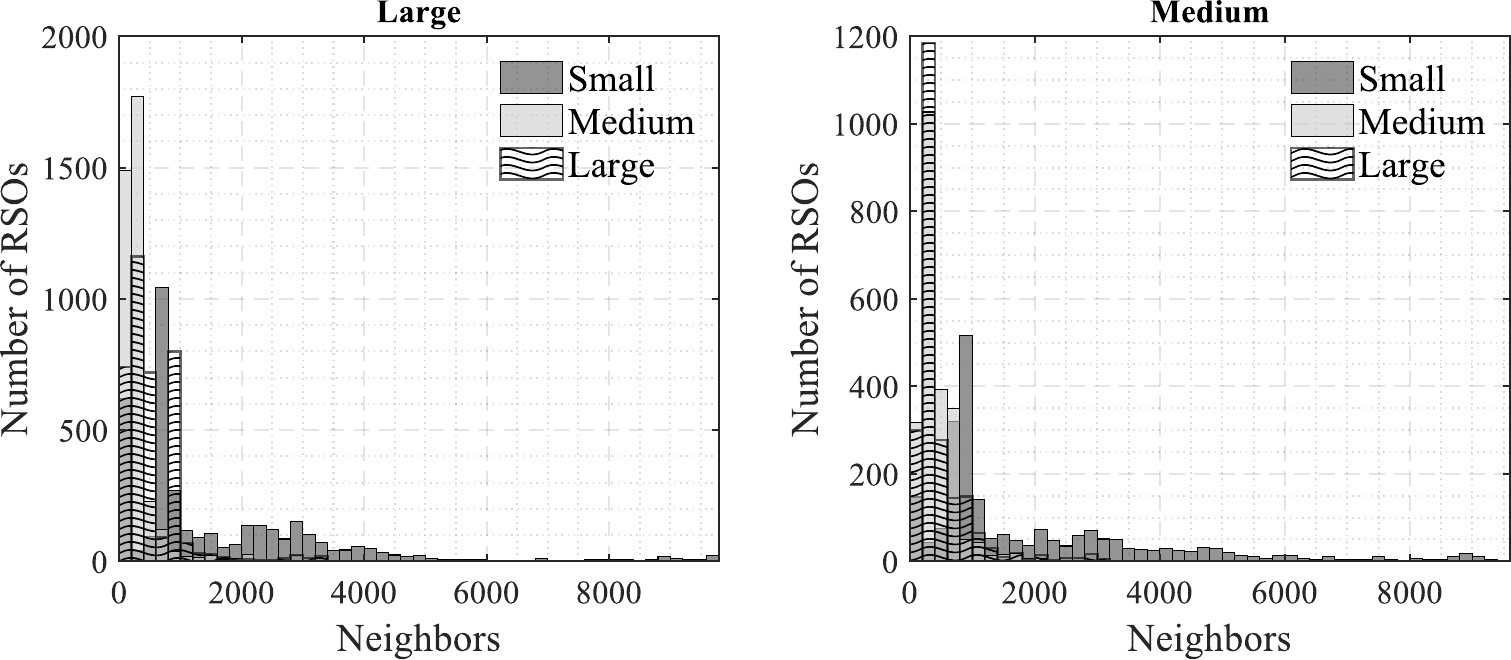}
\caption{Neighbors histograms of large (left) and medium (right) main objects
in 2021.}
\label{hist_21} 
\end{figure}

\begin{figure}[hbt!]
\centering \includegraphics[width=1\textwidth]{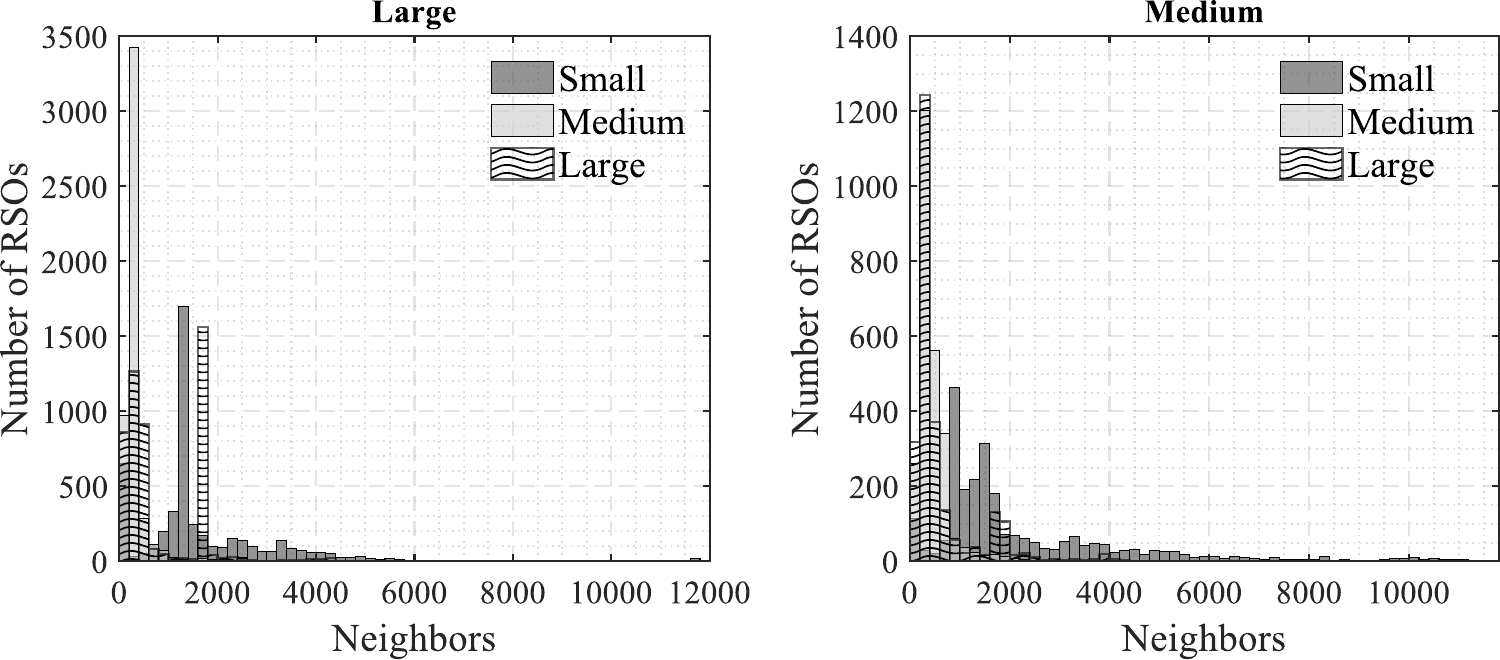}
\caption{Neighbors histograms of large (left) and medium (right) main objects
in 2022.}
\label{hist_22} 
\end{figure}

Figures~\ref{hist_05}-~\ref{hist_22} present histograms of neighborhood
sizes for large and medium objects at five-year intervals from 2005
to 2022. As the years progress, the neighborhoods become larger, with
the maximum size increasing from 3,000 to 12,000. Small object neighborhoods
experienced the most growth, while neighborhoods of large and medium
objects increased slightly, with most having fewer than 1,000 neighbors.
Notably, there was a considerable increase in large-large object pairs
in the last two years, with a sudden rise in neighborhoods containing
1,000-2,000 objects, primarily attributed to the deployment of new
megaconstellations with generally large satellites. Finally, Table
\ref{tab:starlink} presents the neighborhood sizes for two Starlink
satellites in 2022, one at an altitude of approximately 550 km (belonging
to the group 1 Starlink shell) and the other at around 570 km (belonging
to the group 2 shell). Interestingly, while two satellites have similar
number of medium- and small-size neighbors the former interacts with
a much higher number of large-size RSOs, which turn out to belong
to the same group-1 Starlink constellation. Because the two groups
do not overlap radially and group 2 just started to be deployed at
that epoch (January 2022) the large neighbors of the latter satellite
are actually non-Starlink satellites. In fact, it has been verified
that Starlink satellites correspond to the highest peak in the neighbor
histogram on the left hand side of Fig.~\ref{hist_22}. 

\begin{table}[hbt!]
\fontsize{10}{10}\selectfont \caption{Number of neighbors for STARLINK satellites.}
\label{tab:starlink} \centering %
\begin{tabular}{lccccc}
\hline 
NORAD  & Large  & Medium  & Small  & $h_{min}[km]$  & $h_{max}[km]$\tabularnewline
\hline 
44768  & 1,732  & 308  & 1,281  & 538.63  & 553.63\tabularnewline
49134  & 236  & 207  & 1,259  & 566.78  & 574.62\tabularnewline
\hline 
\end{tabular}
\end{table}

\section{Conclusions\label{sec:conclusions}}

This work introduces the concept of short-term space occupancy, a
generalization of the space occupancy framework that improves accuracy
for applications with short time horizons, such as space situational
awareness and conjunction assessment. The short-term SO model is based
on a zonal problem formulation and enables the precise estimation
of an object-occupied altitude range over a given time interval by
solving a quartic equation.

Building upon this theoretical foundation, the SO-filter is proposed.
It is a novel conjunction filter that leverages the short-term SO
model to more accurately identify object pairs with overlapping radial
ranges while accounting for the effects of zonal harmonics. The performance
of the SO-filter is compared to the classical AP-filter and two alternative
SO-filter implementations using a large dataset of space objects and
a high-fidelity propagator.

The results demonstrate that the SO-filter substantially outperforms
the AP-filter in terms of both false positive and false negative rates.
Without applying correction buffers, the SO-filter reduces false positives
by two orders of magnitude and yields eight times fewer false negatives
compared to the AP-filter. When correction buffers are introduced
to eliminate false negatives, the SO-filter requires a buffer three
times smaller than the AP-filter, resulting in a tenfold reduction
in false positives. The performance of the SO-filter is comparable
to the more computationally expensive exact SO-filter implementation,
making it an attractive option for efficient and accurate conjunction
screening.

To showcase the practical utility of the short-term SO model and SO-filter,
an analysis of the evolution of the space population from 2005 to
the present is conducted. Applying these tools to historical space
object data reveals the significant growth in the number of objects
and potential collision pairs over the past two decades. Interestingly,
the analysis also highlights improvements in space traffic management
during this period, as evidenced by the relatively constant number
of object pairs sharing space occupancy between 2010 and 2021, despite
the overall population growth.

However, the sharp increase in shared space occupancy pairs observed
in the last year, primarily due to the deployment of large megaconstellations,
underscores the ongoing challenges in ensuring the safety and sustainability
of the orbital environment. As the space population continues to expand,
the short-term SO model and SO-filter will become increasingly valuable
tools for effective space traffic management and collision risk mitigation.

Future work includes the expansion of these concepts to the classical
orbit path and time filters by continuing to exploit the solutions
of the zonal-perturbed two-body problem. In addition, future extensions
could explore the integration of these methods into existing space
traffic management systems and investigate their potential applications
in other areas, such as mission planning and space debris remediation.

\section*{Acknowledgments}

The authors acknowledge support by grant TED2021-132099B-C33 funded
by MCIN/ AEI/ 10.13039 /501100011033 and by ``European Union NextGenerationEU/PRTR.''
Additional support was provided by MINECO/AEI and FEDER/EU under Project
PID2020-112576GB-C21. Ana S. Rivero was funded by a FPU grant from
the Spanish Ministry of Universities.

The authors thank Prof. Giulio Baù from the University of Pisa for
revising and correcting the derivations of Appendix A.

\bibliography{references}


\section*{Appendix A: Global bounds of the short-term SO \label{Ap:cri_pt}}

\textcolor{black}{This appendix provides a detailed examination of
the mathematical derivation required to identify critical points,
essential for determining the extremal values of the function $r$
within the short-term SO problem, $\left(\hat{\theta}_{i}^{*},\hat{\beta}_{i}^{*}\right)$.
It also includes an exhaustive analysis to determine the type of each
critical point (maximum, minimum, or saddle point) required to ascertain
the global maximum and minimum $\left(\hat{\theta}_{max}^{*},\hat{\beta}_{max}^{*}\right)$
and $\left(\hat{\theta}_{min}^{*},\hat{\beta}_{min}^{*}\right)$,
as well as local extrema, $\left(\hat{\theta}_{Lmax}^{*},\hat{\beta}_{Lmax}^{*}\right)$
and $\left(\hat{\theta}_{Lmin}^{*},\hat{\beta}_{Lmin}^{*}\right)$.}

The analysis primarily centers on computing the first and second derivatives
with respect to both variables, elucidating the fundamental principles
underpinning the computation of critical points. Specifically, the
first derivatives are detailed in Eqs. (\ref{eqn-drdbeta}) and (\ref{eqn-drdtheta}),
while the second derivatives are presented in Eqs. (\ref{eqn-d2rdbeta}),
(\ref{eqn-d2rdbetadtheta}), and (\ref{eqn-d2rdtheta}).

\begin{eqnarray}
\frac{\partial\hat{r}}{\partial\beta} & = & -\hat{a}e_{p}\sin(\hat{\theta}-\beta),\label{eqn-drdbeta}\\
\frac{\partial\hat{r}}{\partial\hat{\theta}} & = & \hat{a}\left(e_{p}\sin(\hat{\theta}-\beta)-e_{f}\cos\hat{\theta}\right)-\frac{J_{2}\sin^{2}\hat{i}}{2\hat{a}}\sin(2\hat{\theta}),\label{eqn-drdtheta}\\
\frac{\partial^{2}\hat{r}}{\partial\beta^{2}} & = & \hat{a}e_{p}\cos(\hat{\theta}-\beta),\label{eqn-d2rdbeta}\\
\frac{\partial^{2}\hat{r}}{\partial\beta\partial\hat{\theta}} & = & -\hat{a}e_{p}\cos(\hat{\theta}-\beta),\label{eqn-d2rdbetadtheta}\\
\frac{\partial^{2}\hat{r}}{\partial\hat{\theta}^{2}} & = & \hat{a}\left(e_{p}\cos(\hat{\theta}-\beta)+e_{f}\sin\hat{\theta}\right)-\frac{J_{2}\sin^{2}\hat{i}}{\hat{a}}\cos(2\hat{\theta}).\label{eqn-d2rdtheta}
\end{eqnarray}

Thus, to find $\hat{\theta}_{i}^{*}$ and $\beta_{i}^{*}$, which
represent the candidates for the extrema, one needs to solve: 
\begin{eqnarray}
0 & = & -\hat{a}e_{p}\sin(\hat{\theta}_{i}^{*}-\beta_{i}^{*}),\label{eqn-drdbeta0}\\
0 & = & \hat{a}\left(e_{p}\sin(\hat{\theta}_{i}^{*}-\beta_{i}^{*})-e_{f}\cos\hat{\theta}_{i}^{*}\right)-\frac{J_{2}\sin^{2}\hat{i}}{2\hat{a}}\sin(2\hat{\theta}_{i}^{*}).\label{eqn-drdtheta0}
\end{eqnarray}

Initially, Eq. (\ref{eqn-drdbeta0}) yields two possibilities: $\hat{\theta}_{i}^{*}=\beta_{i}^{*}$
or $\hat{\theta}_{i}^{*}=\pi+\beta_{i}^{*}$. Substituting these values
into Eq. (\ref{eqn-drdtheta0}) results in the following equation:
\begin{equation}
0=-\hat{a}e_{f}\cos\hat{\theta}_{i}^{*}-\frac{J_{2}\sin^{2}\hat{i}}{2\hat{a}}\sin(2\hat{\theta}_{i}^{*})=-\cos(\hat{\theta}_{i}^{*})\left[\hat{a}e_{f}\textcolor{red}{{\color{black}+}}\frac{J_{2}\sin^{2}\hat{i}}{\hat{a}}\sin(\hat{\theta}_{i}^{*})\right],\label{eqn-othersolution}
\end{equation}
where the expansion of the sine of a double angle has been taken into
account. The possibility $\cos(\hat{\theta}_{i}^{*})=0$ is then considered,
leading to $\hat{\theta}_{i}^{*}=\pi/2$ or $\hat{\theta}_{i}^{*}=-\pi/2$.
This consideration results in four critical points: \textcolor{black}{{} 
\begin{equation}
(\hat{\theta}_{i}^{*},\beta_{i}^{*})=\left\{ (\pi/2,\pi/2),(-\pi/2,\pi/2),(-\pi/2,-\pi/2),(\pi/2,-\pi/2)\right\} ,\label{eq:cri_points1}
\end{equation}
}

\textcolor{black}{Additionally, there exist other critical points
determined by the values of $\hat{\theta}_{i}^{*}$ such that $\hat{a}e_{f}-\frac{J_{2}\sin^{2}\hat{i}}{\hat{a}}\sin(\hat{\theta}_{i}^{*})=0$.
When $\left|\frac{\hat{a}^{2}e_{f}}{J_{2}k}\right|>1$, there are
no solutions of this kind; however, whether $\left|\frac{\hat{a}^{2}e_{f}}{J_{2}k}\right|<1$,
there exist four additional critical points: 
\begin{equation}
\begin{split}(\hat{\theta}_{i}^{*},\beta_{i}^{*}) & =\left\{ \left(-\arcsin\left(\frac{\hat{a}^{2}e_{f}}{J_{2}\sin^{2}\hat{i}}\right),-\arcsin\left(\frac{\hat{a}^{2}e_{f}}{J_{2}\sin^{2}\hat{i}}\right)\right),\left(\pi-\arcsin\left(\frac{\hat{a}^{2}e_{f}}{J_{2}\sin^{2}\hat{i}}\right),-\arcsin\left(\frac{\hat{a}^{2}e_{f}}{J_{2}\sin^{2}\hat{i}}\right)\right),\right.\\
 & \left.\left(-\pi+\arcsin\left(\frac{\hat{a}^{2}e_{f}}{J_{2}\sin^{2}\hat{i}}\right),-\pi+\arcsin\left(\frac{\hat{a}^{2}e_{f}}{J_{2}\sin^{2}\hat{i}}\right)\right),\left(\arcsin\left(\frac{\hat{a}^{2}e_{f}}{J_{2}\sin^{2}\hat{i}}\right),-\pi+\arcsin\left(\frac{\hat{a}^{2}e_{f}}{J_{2}\sin^{2}\hat{i}}\right)\right)\right\} .
\end{split}
\label{eq:cri_points2}
\end{equation}
}

\textcolor{black}{Computing now the Hessian $H$, } 
\begin{equation}
H(\hat{\theta},\beta)=\left[\begin{array}{cc}
\hat{a}e_{p}\cos(\hat{\theta}-\beta) & -\hat{a}e_{p}\cos(\hat{\theta}-\beta)\\
-\hat{a}e_{p}\cos(\hat{\theta}-\beta) & \hat{a}\left(e_{p}\cos(\hat{\theta}-\beta)+e_{f}\sin\hat{\theta}\right)-\frac{J_{2}k}{\hat{a}}\cos(2\hat{\theta})
\end{array}\right].
\end{equation}

One has to remember that a critical point would be a saddle point
if there is no sign definiteness. To have sign definiteness, it is
required that the determinant of the Hessian is positive. Now, the
determinant can be obtained as

\begin{equation}
\mathrm{det}(H)=\hat{a}e_{p}\cos(\hat{\theta}-\beta)\left(\hat{a}e_{f}\sin\hat{\theta}-\frac{J_{2}\sin^{2}\hat{i}}{\hat{a}}\cos(2\hat{\theta})\right).
\end{equation}
At the critical points, the situation unfolds as follows:

\textcolor{black}{{} 
\begin{eqnarray}
H(\pi/2,\pi/2) & = & \left[\begin{array}{cc}
\hat{a}e_{p} & -\hat{a}e_{p}\\
-\hat{a}e_{p} & \hat{a}\left(e_{p}+e_{f}\right)+\frac{J_{2}\sin^{2}\hat{i}}{\hat{a}}
\end{array}\right],\quad\mathrm{det}(H)=\hat{a}^{2}e_{p}e_{f}+e_{p}J_{2}\sin^{2}\hat{i},
\end{eqnarray}
\begin{eqnarray}
H(\pi/2,-\pi/2) & = & \left[\begin{array}{cc}
-\hat{a}e_{p} & \hat{a}e_{p}\\
\hat{a}e_{p} & \hat{a}\left(-e_{p}+e_{f}\right)+\frac{J_{2}\sin^{2}\hat{i}}{\hat{a}}
\end{array}\right],\quad\mathrm{det}(H)=-\hat{a}^{2}e_{p}e_{f}-e_{p}J_{2}\sin^{2}\hat{i},
\end{eqnarray}
\begin{eqnarray}
H(-\pi/2,-\pi/2) & = & \left[\begin{array}{cc}
\hat{a}e_{p} & -\hat{a}e_{p}\\
-\hat{a}e_{p} & \hat{a}\left(e_{p}-e_{f}\right)+\frac{J_{2}\sin^{2}\hat{i}}{\hat{a}}
\end{array}\right],\quad\mathrm{det}(H)=-\hat{a}^{2}e_{p}e_{f}+e_{p}J_{2}\sin^{2}\hat{i},
\end{eqnarray}
\begin{eqnarray}
H(-\pi/2,\pi/2) & = & \left[\begin{array}{cc}
-\hat{a}e_{p} & \hat{a}e_{p}\\
\hat{a}e_{p} & \hat{a}\left(-e_{p}-e_{f}\right)+\frac{J_{2}\sin^{2}\hat{i}}{\hat{a}}
\end{array}\right],\quad\mathrm{det}(H)=\hat{a}^{2}e_{p}e_{f}-e_{p}J_{2}\sin^{2}\hat{i},
\end{eqnarray}
\begin{eqnarray}
H(\hat{\theta}_{5}^{*},\beta_{5}^{*}) & = & \left[\begin{array}{cc}
\hat{a}e_{p} & -\hat{a}e_{p}\\
-\hat{a}e_{p} & \hat{a}e_{p}-\frac{J_{2}^{2}\sin^{4}\hat{i}-\hat{a}^{4}e_{f}^{2}}{\hat{a}J_{2}\sin^{2}\hat{i}}
\end{array}\right],\quad\mathrm{det}(H)=\frac{e_{p}}{J_{2}\sin^{2}\hat{i}}\left(\hat{a}^{4}e_{f}^{2}-J_{2}^{2}\sin^{4}\hat{i}\right),
\end{eqnarray}
\begin{eqnarray}
H(\hat{\theta}_{6}^{*},\beta_{6}^{*}) & = & \left[\begin{array}{cc}
-\hat{a}e_{p} & \hat{a}e_{p}\\
\hat{a}e_{p} & -\hat{a}e_{p}-\frac{J_{2}^{2}\sin^{4}\hat{i}-\hat{a}^{4}e_{f}^{2}}{\hat{a}J_{2}\sin^{2}\hat{i}}
\end{array}\right],\quad\mathrm{det}(H)=-\frac{e_{p}}{J_{2}\sin^{2}\hat{i}}\left(\hat{a}^{4}e_{f}^{2}-J_{2}^{2}\sin^{4}\hat{i}\right),
\end{eqnarray}
\begin{eqnarray}
H(\hat{\theta}_{7}^{*},\beta_{7}^{*}) & = & \left[\begin{array}{cc}
\hat{a}e_{p} & -\hat{a}e_{p}\\
-\hat{a}e_{p} & \hat{a}e_{p}-\frac{J_{2}^{2}\sin^{4}\hat{i}-\hat{a}^{4}e_{f}^{2}}{\hat{a}J_{2}\sin^{2}\hat{i}}
\end{array}\right],\quad\mathrm{det}(H)=\frac{e_{p}}{J_{2}\sin^{2}\hat{i}}\left(\hat{a}^{4}e_{f}^{2}-J_{2}^{2}\sin^{4}\hat{i}\right),
\end{eqnarray}
\begin{eqnarray}
H(\hat{\theta}_{8}^{*},\beta_{8}^{*}) & = & \left[\begin{array}{cc}
-\hat{a}e_{p} & \hat{a}e_{p}\\
\hat{a}e_{p} & -\hat{a}e_{p}-\frac{J_{2}^{2}\sin^{4}\hat{i}-\hat{a}^{4}e_{f}^{2}}{\hat{a}J_{2}\sin^{2}\hat{i}}
\end{array}\right],\quad\mathrm{det}(H)=-\frac{e_{p}}{J_{2}\sin^{2}\hat{i}}\left(\hat{a}^{4}e_{f}^{2}-J_{2}^{2}\sin^{4}\hat{i}\right).
\end{eqnarray}
}

\textcolor{black}{An analysis of the type of critical points will
be conducted, distinguishing between maxima, minima, and saddle points.
In cases where $\left|\frac{\hat{a}^{2}e_{f}}{J_{2}k}\right|>1$,
only the initial four critical points are relevant, with implications
dependent on the sign of $e_{f}$. Typically positive for geocentric
orbits, the value of $e_{f}$ may turn negative very near to critical
inclinations. Therefore, both scenarios should be analyzed.}

\textcolor{black}{When the frozen eccentricity is positive, the points
$(\hat{\theta}_{i}^{*},\beta_{i}^{*})={(\pi/2,\pi/2),(-\pi/2,\pi/2)}$
are the only ones that do not lead to a saddle point. Additionally,
analysis of the sign of the element (1,1) of the Hessian matrix indicates
that the maximum occurs at $\hat{\theta}_{max}^{*}=-\pi/2$, and the
minimum at $\hat{\theta}_{min}^{*}=\pi/2$.}

\textcolor{black}{Conversely, in the event of a negative frozen eccentricity,
the determinant of the Hessian matrix reverses polarity. Consequently,
in this scenario, $(\hat{\theta}_{i}^{*},\beta_{i}^{*})={(\pi/2,-\pi/2),(-\pi/2,-\pi/2)}$
are the points where the extrema are reached. Considering the sign
of the element (1,1) of the Hessian matrix, one can deduce that the
maximum occurs at $\hat{\theta}_{max}^{*}=\pi/2$, while the minimum
is obtained at $\hat{\theta}_{min}^{*}=-\pi/2$.}

\textcolor{black}{When $\left|\frac{\hat{a}^{2}e_{f}}{J_{2}k}\right|<1$,
all eight positions become critical points, independent of the sign
of $e_{f}$. Among the first four candidates, $(\hat{\theta}_{Lmin}^{*},\beta_{Lmin}^{*})={(\pi/2,\pi/2),(-\pi/2,-\pi/2)}$
are the only points that do not lead to a saddle point. Based on the
sign of the element (1,1) of the Hessian matrix, the conclusion is
that both of them are minima. Considering the other four critical
points, $(\hat{\theta}_{Lmax}^{*},\beta_{Lmax}^{*})={(\hat{\theta}_{6}^{*},\beta_{6}^{*}),(\hat{\theta}_{8}^{*},\beta_{8}^{*})}$
are the only ones that do not lead to a saddle point. Further examination
of the sign of the element (1,1) leads to the conclusion that both
represent maxima.}

Note that in cases where only the first four positions are critical
points, the function has only one maximum and one minimum, which corresponds
to the globals, representing the solution of the long-term SO problem
(evaluating the function $r$ at these points yields the expressions
of Eq.\eqref{eq:r_min_LONG} and \eqref{eq:r_max_LONG} ). Conversely,
in instances where all eight positions become critical points, there
are two maxima and two minima. Evaluating $r$ at each position yields
the global maximum,\textcolor{black}{{} $(\hat{\theta}_{max}^{*},\beta_{max}^{*})$},
and minimum,\textcolor{black}{{} $(\hat{\theta}_{min}^{*},\beta_{min}^{*})$},
representing the solution of the long-term SO\footnote{Note that this maximum solution does not correspond to the one obtained
in \cite{SO}, as this work does not account for these particular
cases. These cases stem from the inclusion of additional harmonics
in the calculation of the frozen eccentricity.}. However, the other extrema, which are local,\textcolor{black}{{}
$(\hat{\theta}_{Lmax}^{*},\beta_{Lmax}^{*})$},\textcolor{black}{{}
$(\hat{\theta}_{Lmin}^{*},\beta_{Lmin}^{*})$}, must also be considered
for the analysis of the short-term SO problem.

\section*{Appendix B: Osculating to mean orbital elements conversions \label{Ap:osc_mean}}

The algorithm employed to convert the osculating orbit elements into
mean orbit elements is outlined in this appendix. It is based on the
theory developed by Kozai \cite{Kozai} and Lyddane \cite{Lyddane}.

This mapping translates any osculating (instantaneous) orbital elements
into mean (orbit averaged, with short period motion removed) orbital
element equivalent values. It is important to take into account that
only first order $J_{2}$ terms are retained in this algorithm, thus
small errors of order $J_{2}^{2}$ are to be expected. As it is explained
in Appendix G of Schaub \cite{Schaub}, the forward and inverse mapping
functions between the mean and osculating orbit elements only differ
by a sign, because a first-order truncation is performed for the infinite
power series solution. In this appendix, the equations are written
with the signs corresponding to the change from osculating to mean
elements (to compute the osculating elements from the mean ones, it
is enough to switch the signs of the short-periodic terms).

Let the original osculating orbit elements be given by $(a,e,i,\Omega,\omega,M)$
and the transformed mean orbit elements be given through $(\hat{a},\hat{e},\hat{i},\hat{\Omega},\hat{\omega},\hat{M})$.
The true anomaly, $\nu$, is computed using Kepler's equation, given
in Eq.~\eqref{Kepler}, and the eccentric anomaly, E, is related
with $\nu$ through Eq.~\eqref{nu}.

\begin{equation}
M=E-e\sin E,\label{Kepler}
\end{equation}

\begin{equation}
\nu=2\tan^{-1}\left(\sqrt{\frac{1+e}{1-e}}\tan\left(\frac{E}{2}\right)\right).\label{nu}
\end{equation}

Following Kozai, the short-periodic term for the orbital elements
of the zonal problem are computed by Eqs.~\eqref{a_sp}-\eqref{M_sp}
given next. It is important to notice that the mean values of short-periodic
perturbations are not zero, except those of semimajor axis. Therefore,
in these equations, their mean values with respect to the mean anomaly
are subtracted. This subtraction is the last term of Eqs.~\eqref{e_sp}-\eqref{M_sp}.

\textit{Semimajor axis:}

\begin{equation}
a_{sp}=\frac{J_{2}}{2a}\left[\left(2-3\kappa\right)\left(\frac{a^{3}}{r^{3}}-\lambda^{-3}\right)+3\kappa\frac{a^{3}}{r^{3}}c_{2,2}\right].\label{a_sp}
\end{equation}

\textit{Eccentricity:}

\begin{equation}
\begin{split}e_{sp}=\frac{\lambda^{2}}{2e}\frac{3J_{2}}{a^{2}}\left[\frac{1}{3}\left(1-\frac{3}{2}\kappa\right)\left(\frac{a^{3}}{r^{3}}-\lambda^{-3}\right)+\frac{1}{2}\frac{a^{3}}{r^{3}}\kappa c_{2,2}\right]\\
\\
-\frac{3J_{2}\kappa}{4ea^{2}\lambda^{2}}\left[c_{2,2}+ec_{1,2}+\frac{1}{3}ec_{3,2}\right]-\frac{J_{2}\kappa e(2\lambda+1)\cos(2\omega)}{4a^{2}\lambda^{2}(\lambda+1)^{2}}.
\end{split}
\label{e_sp}
\end{equation}

\textit{Inclination:}

\begin{equation}
i_{sp}=\frac{J_{2}}{8a^{2}\lambda^{4}}\sin2i\left(3c_{2,2}+3ec_{1,2}+ec_{3,2}\right)-\frac{J_{2}\sin2i(2\lambda^{2}-\lambda-1)\cos(2\omega)}{8a^{2}\lambda^{4}(\lambda+1)}.
\end{equation}

\textit{Argument of pericenter: } 
\begin{equation}
\begin{split}\omega_{sp}=\dfrac{3J_{2}}{2a^{2}\lambda^{4}}\left\{ \dfrac{4-5\kappa}{2}\left(\nu-M+e\,s_{1,0}\right)+\dfrac{5\kappa-2}{4}\left(s_{2,2}+es_{1,2}+\dfrac{e}{3}s_{3,2}\right)\right.\\
\\
+\dfrac{1}{e}\left[\left(\dfrac{2-3\kappa}{2}\right)\left[\left(1-\dfrac{e^{2}}{4}\right)s_{1,0}+\dfrac{e}{2}s_{2,0}+\dfrac{e^{2}}{12}s_{3,0}\right]\right.\\
\\
-\kappa\left(\dfrac{1}{4}\left(1+\dfrac{5}{4}e^{2}\right)s_{1,2}-\dfrac{e^{2}}{16}s_{1,-2}\right.\left.\left.\left.-\dfrac{7}{12}\left(1-\dfrac{e^{2}}{28}\right)s_{3,2}-\dfrac{3}{8}es_{4,2}-\dfrac{e^{2}}{16}s_{5,2}\right)\right]\right\} \\
\\
-\dfrac{3J_{2}}{2a^{2}\lambda^{4}}\left[\dfrac{\kappa}{8}+\dfrac{\left(1+2\lambda\right)\left(2\kappa\lambda^{2}-\lambda^{2}-\kappa+1\right)}{6\left(\lambda+1\right)^{2}}\right]\sin\left(2\omega\right).
\end{split}
\end{equation}

\textit{Longitude of the ascending node:}

\begin{equation}
\begin{split}\Omega_{sp}=-\frac{3J_{2}}{2a^{2}\lambda^{4}}\cos i\left[\nu-M+e\sin\nu-\frac{1}{2}s_{2,2}-\frac{1}{2}es_{1,2}-\frac{1}{6}es_{3,2}\right]\\
\\
-\frac{J_{2}\cos i(2\lambda^{2}-\lambda-1)\sin(2\omega)}{4a^{2}\lambda^{4}(\lambda+1)}.
\end{split}
\end{equation}

\textit{Mean anomaly:}

\begin{equation}
\begin{split}eM_{sp}=\frac{3J_{2}}{2a^{2}\lambda^{3}}\left\lbrace -\left(1-\frac{3}{2}\kappa\right)\left[\left(1-\frac{e^{2}}{4}\right)\sin\nu+\frac{e}{2}\sin2\nu+\frac{e^{2}}{12}\sin3\nu\right]\right.\\
\\
+\kappa\left[\frac{1}{4}\left(1+\frac{5}{4}e^{2}\right)\sin(\nu+2\omega)-\frac{e^{2}}{16}\sin(\nu-2\omega)-\frac{7}{12}\left(1-\frac{e^{2}}{28}\right)\sin(3\nu+2\omega)\right.\\
\\
\left.\left.-\frac{3}{8}e\sin(4\nu+2\omega)-\frac{e^{2}}{16}\sin(5\nu+2\omega)\right]\right\rbrace +\frac{eJ_{2}\kappa(4\lambda^{3}-\lambda^{2}-18\lambda-9)\sin2\omega}{16a^{2}\lambda^{3}(\lambda+1)^{2}}.
\end{split}
\label{M_sp}
\end{equation}

The following terms were used in Eqs.~\eqref{a_sp}-\eqref{M_sp}:

\begin{equation}
\centering\begin{split}\lambda=\sqrt{1-e^{2}},\;\kappa=\sin^{2}i,\;r=\frac{a(1-e^{2})}{1+e\cos\nu},\\
\\
s_{1,0}=\sin\nu,\;s_{2,0}=\sin2\nu,\;s_{3,0}=\sin3\nu,\;s_{1,2}=\sin(\nu+2\omega),\\
\\
s_{1,-2}=\sin(\nu-2\omega),\;s_{2,2}=\sin(2\nu+2\omega),\;s_{3,2}=\sin(3\nu+2\omega),\\
\\
s_{4,2}=\sin(4\nu+2\omega),\;s_{5,2}=\sin(5\nu+2\omega),\;c_{1,2}=\cos(\nu+2\omega),\\
\\
c_{2,2}=\cos(2\nu+2\omega),\;c_{3,2}=\cos(3\nu+2\omega).
\end{split}
\end{equation}

These expressions, with the exception of the semimajor axis, are well
known to contain singularities both near-circular and/or equatorial
orbits. In order to obtain a more robust mapping near these conditions,
Lyddane's theory is applied. To provide numerically stable expressions
for the mean anomaly and eccentricity short-periodic components, $\varsigma$
and $\iota$ are defined next by Eqs.~\eqref{vsigma}~and~\eqref{iota}.
Similarly, for the right ascension of the ascending node and the inclination,
$\rho$ and $\phi$ are determined by Eqs.~\eqref{rho}~and~\eqref{phi}.

\begin{eqnarray}
\varsigma & = & (e-e_{sp})\cos M+eM_{sp}\sin M\approx(e-e_{sp})\cos(M-M_{sp}),\label{vsigma}\\
\iota & = & (e-e_{sp})\sin M-eM_{sp}\cos M\approx(e-e_{sp})\sin(M-M_{sp}),\label{iota}\\
\rho & = & \left(\sin\frac{i}{2}-\frac{i_{sp}}{2}\cos\frac{i}{2}\right)\cos\Omega+\sin\frac{i}{2}\sin\Omega\Omega_{sp}\approx\sin\left(\frac{i-i_{sp}}{2}\right)\cos\left(\Omega-\Omega_{sp}\right),\label{rho}\\
\phi & = & \left(\sin\frac{i}{2}-\frac{i_{sp}}{2}\cos\frac{i}{2}\right)\sin\Omega+\sin\frac{i}{2}\cos\Omega\Omega_{sp}\approx\sin\left(\frac{i-i_{sp}}{2}\right)\sin\left(\Omega-\Omega_{sp}\right).\label{phi}
\end{eqnarray}

Once these variables are defined, numerically stable expressions for
the five orbital elements can be computed with Eqs~\eqref{M_st}-\eqref{omega_st},
whereas the mean semimajor axis is computed by Eq.~\eqref{a_m}.

\begin{eqnarray}
\hat{a} & = & a-a_{sp},\label{a_m}\\
\hat{M} & = & \tan^{-1}\left(\frac{\iota}{\varsigma}\right),\label{M_st}\\
\hat{e} & = & \sqrt{\iota^{2}+\varsigma^{2}},\\
\hat{\Omega} & = & \tan^{-1}\left(\frac{\phi}{\rho}\right),\\
\hat{i} & = & 2\sin^{-1}\left(\sqrt{\phi^{2}+\rho^{2}}\right),\\
\hat{\omega} & = & (M-M_{sp})+(\omega-\omega_{sp})+(\Omega-\Omega_{sp})-\hat{M}-\hat{\Omega}.\label{omega_st}
\end{eqnarray}

\end{document}